\documentclass[12pt]{article}

\usepackage{graphicx}

\usepackage[square,comma,numbers,sort&compress]{natbib}
\usepackage[left=20mm,right=20mm,top=20mm,bottom=20mm]{geometry}
\usepackage{bm}
\usepackage{comment}

\usepackage{epsf}
\usepackage{color}
\usepackage{amsmath}
\usepackage{amssymb}
\usepackage{latexsym}
\usepackage{braket}
\usepackage{slashed}
\usepackage{float}
\usepackage{cases}
\usepackage{bm}
\usepackage{arydshln}

\newcommand{\B}{\mathcal{B}}

\newcommand{\V}{\mathcal{V}}

\newcommand{\al}[1]{\begin{align}#1\end{align}}

\newcommand{\bp}{\begin{pmatrix}}
\newcommand{\ep}{\end{pmatrix}}
\newcommand{\bb}{\begin{bmatrix}}
\newcommand{\eb}{\end{bmatrix}}

\def\lsim{\mathrel{\rlap{\lower4pt\hbox{\hskip1pt$\sim$}}
    \raise1pt\hbox{$<$}}}         
\def\gsim{\mathrel{\rlap{\lower4pt\hbox{\hskip1pt$\sim$}}
    \raise1pt\hbox{$>$}}}         

\definecolor{mygrn}{rgb}{0,0.6,0}

\usepackage[normalem]{ulem}

\newcommand{\beq}{\begin{equation}}
\newcommand{\eeq}{\end{equation}}
\newcommand{\bea}{\begin{eqnarray}}
\newcommand{\eea}{\end{eqnarray}}

\newcommand{\fulltoday}{\number\day\space \ifcase\month\or
    January\or February\or March\or April\or May\or June\or
    July\or August\or September\or October\or November\or December\fi
    \space\number\year}

 \makeatletter
    
    \@addtoreset{equation}{section}
  \makeatother

\usepackage{calc}
\newcounter{hours}\newcounter{minutes}
\makeatletter                   
\renewcommand*{\thehours}{\two@digits\c@hours}
\renewcommand*{\theminutes}{\two@digits\c@minutes}
\makeatother

\begin{document}

\allowdisplaybreaks[2]

\begin{titlepage}

\renewcommand\thefootnote{\alph{footnote}}
		\mbox{}\hfill CTPU-15-29\\
		\mbox{}\hfill KIAS-P15067\\
		\mbox{}\hfill UT-HET-110\\
\vspace{4mm}
\begin{center}
{\fontsize{22pt}{0pt}\selectfont \bf {
{LHC 750\,GeV Diphoton excess \\[6pt] in a radiative seesaw model}
}} \\
\vspace{8mm}
	{\fontsize{14pt}{0pt}\selectfont \bf
	Shinya Kanemura,\,\footnote{
		E-mail: \tt kanemu@sci.u-toyama.ac.jp
		}}
	{}
	{\fontsize{14pt}{0pt}\selectfont \bf
	Kenji Nishiwaki,\,\footnote{
		E-mail: \tt nishiken@kias.re.kr
		}}
	{}
	{\fontsize{14pt}{0pt}\selectfont \bf
	Hiroshi Okada,\,\footnote{
		E-mail: \tt macokada3hiroshi@gmail.com
		}}
	\\[3pt]
	{\fontsize{14pt}{0pt}\selectfont \bf
	Yuta Orikasa,\,\footnote{
		E-mail: \tt orikasa@kias.re.kr
		}}
	{}
	{\fontsize{14pt}{0pt}\selectfont \bf
	Seong Chan Park,\,\,\footnote{
		E-mail: \tt sc.park@yonsei.ac.kr} }
	{}
	{\fontsize{14pt}{0pt}\selectfont \bf
	Ryoutaro Watanabe\,\,\footnote{
		E-mail: \tt wryou1985@ibs.re.kr} } \\
\vspace{4mm}
	{\fontsize{10pt}{0pt}\selectfont
		${}^{\mathrm{a}}$\,\it Department of Physics, University of Toyama,\\[3pt]
		                     3190 Gofuku, Toyama 930-8555, Japan
		\smallskip\\[3pt]
		${}^{\mathrm{b,d,e}}$\,\it School of Physics, Korea Institute for Advanced Study,\\[3pt]
		                           Seoul 02455, Republic of Korea
		\smallskip\\[3pt]
		${}^{\mathrm{c}}$\,\it Physics Division, National Center for Theoretical Sciences,\\[3pt]
                               National Tsing-Hua University, Hsinchu 30013, Taiwan
		\smallskip\\[3pt]
		${}^{\mathrm{{d}}}$\,\it Department of Physics and Astronomy, Seoul National University,\\[3pt]
		                       Seoul 08826, Republic of Korea
		\smallskip\\[3pt]
		${}^{\mathrm{e}}$\,\it Department of Physics and IPAP, Yonsei University,\\[3pt]
		                       Seoul {03722}, Republic of Korea
		\smallskip\\[3pt]
		${}^{\mathrm{f}}$\,\it Center for Theoretical Physics of the Universe,\\[3pt]
                               Institute for Basic Science (IBS), Daejeon, 34051, Republic of Korea
		\smallskip\\
	}
\vspace{4mm}
{\normalsize \fulltoday}
\vspace{10mm}
\end{center}
\begin{abstract}
{\fontsize{10pt}{16pt}\selectfont{
We investigate a possibility for explaining the recently announced 750\,GeV diphoton excess by the ATLAS and the CMS experiments at the CERN LHC in a model with multiple doubly charged particles, which was originally suggested for explaining tiny neutrino masses through a three-loop effect in a natural way.
The enhanced radiatively generated effective coupling of a new singlet scalar $S$ with diphoton with multiple charged particles in the loop enlarges the production rate of $S$ in $pp\to S+X$ via photon fusion process and also the decay width $\Gamma(S\to \gamma\gamma)$ even without assuming a tree level production mechanism. We provide detailed analysis on the cases with or without allowing the mixing between $S$ and the standard model Higgs doublet.
}
}
\end{abstract}
\end{titlepage}
\renewcommand\thefootnote{\arabic{footnote}}
\setcounter{footnote}{0}

\section{Introduction}

In the mid of December 2015, both of the ATLAS and CMS experiments announced the observation of a new resonance around 750\,GeV as a bump in the diphoton invariant mass spectrum from the run-II data in $\sqrt{s} = 13\,\text{TeV}$~\cite{ATLAS-CONF-2015-081,CMS-PAS-EXO-15-004}.
Their results are based on the accumulated data of $3.2\,\text{fb}^{-1}$ (ATLAS) and $2.6\,\text{fb}^{-1}$ (CMS), and local/global significances are $3.9\sigma/2.3\sigma$ (ATLAS)~\cite{ATLAS-CONF-2015-081} and $2.6\sigma/\lesssim 1.2\sigma$ (CMS)~\cite{CMS-PAS-EXO-15-004}, respectively.
The best-fit values of the invariant mass are $750\,\text{GeV}$ by ATLAS and $760\,\text{GeV}$ by CMS, where the ATLAS also reported the best-fit value of the total width as $45\,\text{GeV}$.

During/after Moriond EW in March 2016, updated results were reported {with the new analysis with the different hypotheses on spin (spin-0 or spin-2) and the width to mass ratio ($\Gamma/m < 1\%$ `narrow width' or $\Gamma/m \sim 6-10\%$ `wide width')}~\cite{ATLAS-CONF-2016-018,CMS-PAS-EXO-16-018}.
Based on the $3.2\,\text{fb}^{-1}$ dataset, the ATLAS group claimed that the largest deviation from the background-only hypothesis was observed near a mass of $750\,\text{GeV}$, which corresponds to a local excess of $3.9\sigma$ for  the spin-0 case of $\Gamma \approx 45\,\text{GeV} \ (\Gamma/m \approx 6\%)$. {However, we note that the preference of the `wide width' compared with `narrow width'  is only minor by $\sim 0.3 \sigma$ significance so that we would take it with cautious attention. In our analysis below, we simply allow both cases with narrow and wide widths.} 
The global significance {is still low $\sim 2.0 \sigma$.}

{On the other hand,} based on the upgraded amount of the data of $3.3\,\text{fb}^{-1}$, the CMS group reported a modest excess of events at $760\,\text{GeV}$ with a local significance of $2.8-2.9\sigma$ depending on the spin hypothesis. The `narrow width' {($\Gamma/m = 1.4 \times 10^{-2}$)} {maximizes the local excess.}
In addition, {the CMS reported} the result of a combined analysis of $8\,\text{TeV}$ and $13\,\text{TeV}$ data, where the largest excess ($3.4 \sigma$) was observed at $750\,\text{GeV}$ for the narrow width ($\Gamma/m = 1.4 \times 10^{-4}$).
The global significances are $<1\sigma$ ($1.6\sigma$) in the $13\,\text{TeV}$ ($8\,\text{TeV} + 13\,\text{TeV}$) analyses, respectively.
No official combined (ATLAS \& CMS) result {has been made so far}. 


Just after the advent of the first announcement, various ways for explaining the 750\,GeV excess have been proposed even within December 2015 in Refs.~\cite{Harigaya:2015ezk,Mambrini:2015wyu,Backovic:2015fnp,Angelescu:2015uiz,Nakai:2015ptz,Knapen:2015dap,Buttazzo:2015txu,Pilaftsis:2015ycr,Franceschini:2015kwy,
DiChiara:2015vdm,Higaki:2015jag,McDermott:2015sck,Ellis:2015oso,Low:2015qep,Bellazzini:2015nxw,Gupta:2015zzs,Petersson:2015mkr,Molinaro:2015cwg,Falkowski:2015swt,
Dutta:2015wqh,Cao:2015pto,Matsuzaki:2015che,Kobakhidze:2015ldh,Martinez:2015kmn,Cox:2015ckc,Becirevic:2015fmu,No:2015bsn,Demidov:2015zqn,Chao:2015ttq,Fichet:2015vvy,Curtin:2015jcv,Bian:2015kjt,Chakrabortty:2015hff,Ahmed:2015uqt,Agrawal:2015dbf,Csaki:2015vek,Aloni:2015mxa,Bai:2015nbs,
Gabrielli:2015dhk,Benbrik:2015fyz,Kim:2015ron,Alves:2015jgx,Megias:2015ory,Carpenter:2015ucu,Bernon:2015abk,
Chao:2015nsm,Arun:2015ubr,Han:2015cty,Chang:2015bzc,Chakraborty:2015jvs,Ding:2015rxx,Han:2015dlp,Han:2015qqj,Luo:2015yio,Chang:2015sdy,Bardhan:2015hcr,Feng:2015wil,Antipin:2015kgh,Wang:2015kuj,Cao:2015twy,Huang:2015evq,Liao:2015tow,Heckman:2015kqk,Dhuria:2015ufo,Bi:2015uqd,Kim:2015ksf,Berthier:2015vbb,Cho:2015nxy,Cline:2015msi,Bauer:2015boy,Chala:2015cev,Barducci:2015gtd,Pelaggi:2015knk,
Boucenna:2015pav,Murphy:2015kag,Hernandez:2015ywg,Dey:2015bur,deBlas:2015hlv,Belyaev:2015hgo,Dev:2015isx,
Huang:2015rkj,Moretti:2015pbj,Patel:2015ulo,Badziak:2015zez,Chakraborty:2015gyj,Cao:2015xjz,Altmannshofer:2015xfo,Cvetic:2015vit,Gu:2015lxj,
Allanach:2015ixl,Davoudiasl:2015cuo,Craig:2015lra,Das:2015enc,Cheung:2015cug,Liu:2015yec,Zhang:2015uuo,Casas:2015blx,Hall:2015xds,
Han:2015yjk,Park:2015ysf,Salvio:2015jgu,Chway:2015lzg,Li:2015jwd,Son:2015vfl,Tang:2015eko,An:2015cgp,Cao:2015apa,Wang:2015omi,Cai:2015hzc,Cao:2015scs,Kim:2015xyn,Gao:2015igz,Chao:2015nac,Bi:2015lcf,Goertz:2015nkp,Anchordoqui:2015jxc,Dev:2015vjd,Bizot:2015qqo,
Hamada:2015skp,Kanemura:2015vcb,Jiang:2015oms}.
The first unofficial interpretation of the excess in terms of the signal strength of a scalar (or a pseudoscalar) resonance $S$, {$pp \to S+X \to \gamma\gamma+X$}, was done immediately after the first announcement in Ref.~\cite{Buttazzo:2015txu} based on the expected and observed exclusion limits in both of the experiments.
The authors claimed,
\al{
\mu^{\text{ATLAS}}_{13\text{TeV}} = \sigma({pp \to S+X})_{13\text{TeV}} \times \B(S \to \gamma\gamma)
&= ({10^{+4}_{-3}})\,\text{fb},
	\label{eq:13TeVATLAS_xsec_target} \\
\mu^{\text{CMS}}_{13\text{TeV}} = \sigma({pp \to S+X})_{13\text{TeV}} \times \B(S \to \gamma\gamma)
&= (5.6 \pm 2.4)\,\text{fb},
	\label{eq:13TeVCMS_xsec_target}
	}
with a Poissonian likelihood function (for the {ATLAS} measurement) and the Gaussian approximation (for the {CMS} measurement), respectively.

On the other hand, both of the ATLAS and CMS groups reported that no significant excess over the standard model (SM) background was observed in their analyses based on the run-I data at $\sqrt{s} = 8\,\text{TeV}$~\cite{Aad:2015mna,CMS-PAS-HIG-14-006}, while a mild upward bump was found in their data around $750\,\text{GeV}$.
In Ref.~\cite{Buttazzo:2015txu}, the signal strengths at $\sqrt{s} = 8\,\text{TeV}$ were extracted by use of the corresponding expected and observed exclusion limits given by the experiments, in the Gaussian approximation, for a narrow-width scalar resonance as
\al{
\mu^{\text{ATLAS}}_{8\text{TeV}} = \sigma({pp \to S+X})_{8\text{TeV}} \times \B(S \to \gamma\gamma)
&= (0.46 \pm 0.85)\,\text{fb},
	\label{eq:8TeVATLAS_xsec_target} \\
\mu^{\text{CMS}}_{8\text{TeV}} = \sigma({pp \to S+X})_{8\text{TeV}} \times \B(S \to \gamma\gamma)
&= (0.63 \pm {0.35})\,\text{fb}.
	\label{eq:8TeVCMS_xsec_target}
}
It is mentioned that when we upgrade the collider energy from $8\,\text{TeV}$ to $13\,\text{TeV}$, a factor $4.7$ enhancement is expected~\cite{Buttazzo:2015txu,8TeV-to-13TeV-factor}, when the resonant particle is produced via gluon fusion, and then the data at $\sqrt{s} = 8\,\text{TeV}$ and $13\,\text{TeV}$ are compatible at around $2\sigma$ confidence level (C.L.).
{Indeed, in the second announcement~\cite{ATLAS-CONF-2016-018}}, the ATLAS group discussed this point based on the reanalyzed $8\,\text{TeV}$ data corresponding to an integrated luminosity of $20\,\text{fb}^{-1}$ with the latest photon energy calibration in the run-I, which is close to the calibration used for the $13\,\text{TeV}$ data.
When $m=750\,\text{GeV}$ and $\Gamma/m = 6\%$, the difference between the $8\,\text{TeV}$ and {$13\,\text{TeV}$ }results corresponds to statistical significances of $1.2\sigma$ ($2.1\sigma$) if gluon-gluon (quark-antiquark) productions are assumed.
These observations would give us a stimulating hint for surveying the structure of physics beyond the SM above the electroweak scale even though the accumulated amount of the data {would not be} enough for detailed discussions and the errors are large at the present stage.

A key point to understand the resonance is the fact that no bump around $750\,\text{GeV}$ has been found in the other final states in both of the $8\,\text{TeV}$ and $13\,\text{TeV}$ data.
If $\B(S \to \gamma\gamma)$ is the same as the $750\,\text{GeV}$ Higgs one, $\B(h \to \gamma\gamma)|_{750\,\text{GeV}\,\text{SM}} = 1.79 \times 10^{-7}$~\cite{Heinemeyer:2013tqa},
we can immediately recognize that such a possibility is inconsistent with the observed results, e.g., in $ZZ$ final state, at $\sqrt{s} = 8\,\text{TeV}$, where {the significant experimental $95\%$ C.L. upper bound on the $ZZ$ channel is $12\,\text{fb}$ by ATLAS~\cite{Aad:2015kna}}
and the branching ratio $\B(h \to ZZ)|_{750\,\text{GeV}\,\text{SM}} = 0.290$~\cite{Heinemeyer:2013tqa}.
In general, the process $S \to \gamma\gamma$ should be loop induced since $S$ has zero electromagnetic charge and then the value of $\B(S \to \gamma\gamma)$ tends to be suppressed because tree level decay branches generate primary components of the total width of {$S$}.
Then, a reasonable setup for explaining the resonance consistently is that all of the decay channels of $S$ are one loop induced, where $S$ would be a gauge singlet under $SU(3)_C$ and $SU(2)_L$ {since a non-singlet gauge assignment leads to tree level gauge interactions, which are not desired in our case.}

An example of this direction is that $S$ is a singlet scalar and it couples to vector-like quarks, which contribute to both of {$p p \to S+X$} and $S \to \gamma\gamma$ via gluon fusion and photon fusion, respectively.
{The possibility of diphoton production solely due to photon fusion is also an open possibility as discussed in {Refs}.~\cite{Fichet:2015vvy,Csaki:2015vek} in the context of the $750\,\text{GeV}$ excess.} {Basic idea is simple: when a model contains {multiple} $SU(2)_L$ singlet particles with large $U(1)_Y$ hypercharges, the magnitude of the photon fusions in the production and decay sequences is largely enhanced.}

In this {paper}, we focus on the radiative {seesaw} models~\cite{Zee:1980ai,Cheng:1980qt,Zee:1985id,Babu:1988ki,Ma:2006km}, especially where neutrino masses are generated at the three loop level~\cite{Krauss:2002px,Aoki:2008av,Aoki:2011zg,Gustafsson:2012vj,Kajiyama:2013lja,Ahriche:2014cda,Ahriche:2014oda,Chen:2014ska,Okada:2014oda,Hatanaka:2014tba,Jin:2015cla,Culjak:2015qja,Geng:2015coa,Ahriche:2015wha,Nishiwaki:2015iqa,Okada:2015hia,Ahriche:2015loa,Ahriche:2015taa}.
In such type of {scenarios}, {multiple} charged scalars are {introduced} for realizing three-loop origin of the neutrino mass, ({distinctively from the models with one or two loops}). We show that when these charged scalars couple to the singlet $S$ {strongly enough}, we can achieve a reasonable amount of the production cross section in $p p \to S+X \to \gamma\gamma +X$ through photon fusion.
Concretely, we start from the three loop model~\cite{Nishiwaki:2015iqa}, and extend the model with 
additional charged scalars to explain the data.\footnote{
Recently, several other works have emerged in this direction~\cite{Nomura:2016fzs,Nomura:2016seu,Okada:2016rav,Nomura:2016rjf,Arbelaez:2016mhg,Ko:2016sxg}.
}

This {paper} is organized as follows.
In Sec.~\ref{sec:model}, {we introduce our model based on the model for three-loop induced neutrino masses.}
In Sec.~\ref{sec:analysis}, we show detail of analysis and numerical results.
In Sec.~\ref{sec:summary}, we are devoted to summary and discussions.

\section{Model
\label{sec:model}}
{
Multiple (doubly) charged particles would induce a large radiative coupling with a singlet scalar $S$ with $\gamma\gamma$ via one-loop diagrams. 
We may find the source from multi-Higgs models or extra dimensions
{\cite{Park:2009cs,Chen:2009gz,Chen:2009zs,Park:2009gp,Kong:2010xk,Kong:2010qd,Csaki:2010az,Nishiwaki:2011vi,Nishiwaki:2011gk,Nishiwaki:2011gm,Kim:2011tq,Datta:2012tv,Flacke:2013pla,Kakuda:2013kba,Flacke:2013nta,Datta:2013yaa,Dohi:2014fqa,Flacke:2014jwa}}
but here we focus on a model for radiative neutrino masses recently suggested by some of the authors~\cite{Nishiwaki:2015iqa} as a benchmark model, which can be extended with a singlet scalar $S$ for the $750$ GeV resonance.
}

\subsection{Review: A model for three-loop induced neutrino mass}

Our strategy is based on the three loop induced radiative neutrino model with a $U(1)$ global symmetry~\cite{Nishiwaki:2015iqa}, where
we introduce three Majorana fermions $N_{R_{1,2,3}}$ and new bosons; one gauge-singlet neutral boson $\Sigma_0$, two
{singly charged} singlet scalars ($h^\pm_1, h^\pm_2$), and one gauge-singlet {doubly charged} boson $k^{\pm\pm}$ to {the} SM.
The particle contents and their charges are shown in {Table}~\ref{tab:1}.

\begin{table}[h]
\begin{tabular}{|c||c|c|c||c|c|c|c|c||c|c|}\hline\hline  
&\multicolumn{3}{c||}{Lepton Fields} & \multicolumn{5}{c||}{Scalar Fields} & \multicolumn{2}{c|}{New Scalar Fields} \\\hline
{Characters} & ~$L_{L_i}$~ & ~$e_{R_i}$~ &~$N_{R_i}$~ & ~$\Phi$~ &
 ~$\Sigma_0$~ & ~$h^+_1$~  & ~$h^{+}_2$~ & ~$k^{++}$~ & ~$j_a^{++}$~ & $S$ \\\hline 
$SU(3)_C$ & $\bm{1}$ & $\bm{1}$ & $\bm{1}$ & $\bm{1}$ & $\bm{1}$ & $\bm{1}$ & $\bm{1}$ & $\bm{1}$ & $\bm{1}$ & $\bm{1}$ \\ \hline
$SU(2)_L$ & $\bm{2}$ & $\bm{1}$&  $\bm{1}$&$\bm{2}$ & $\bm{1}$ &$\bm{1}$  &$\bm{1}$ & $\bm{1}$ & $\bm{1}$ & $\bm{1}$ \\\hline 
$U(1)_Y$ & $-1/2$ & $-1$ & $0$ & $1/2$  & $0$  & $1$  & $1$ & $2$ & $2$ & 0 \\\hline
{$U(1)$} & $0$ & $0$ & $-x$ & $0$  & $2x$   & $0$ & $x$ & $2x$ & $2x$ & $0$  \\\hline
\end{tabular}
\centering
\caption{Contents of lepton and scalar fields
and their charge assignment under $SU(3)_C \times SU(2)_L\times U(1)_Y\times {U(1)}$, {where $U(1)$  is an additional global symmetry and} $x\neq 0$.
The subscripts found in the lepton fields $i \,(=1,2,3)$ indicate generations of the fields.
The bold letters emphasize that these numbers correspond to representations of the Lie groups of the NonAbelian gauge interactions.
The scalar particles shown in the right category (New Scalar Fields) are added {to} the original model proposed in Ref.~\cite{Nishiwaki:2015iqa} to explain the 750\,GeV excess.
}
\label{tab:1}
\end{table}

\begin{figure}[t]
\begin{center}
\includegraphics[scale=0.5]{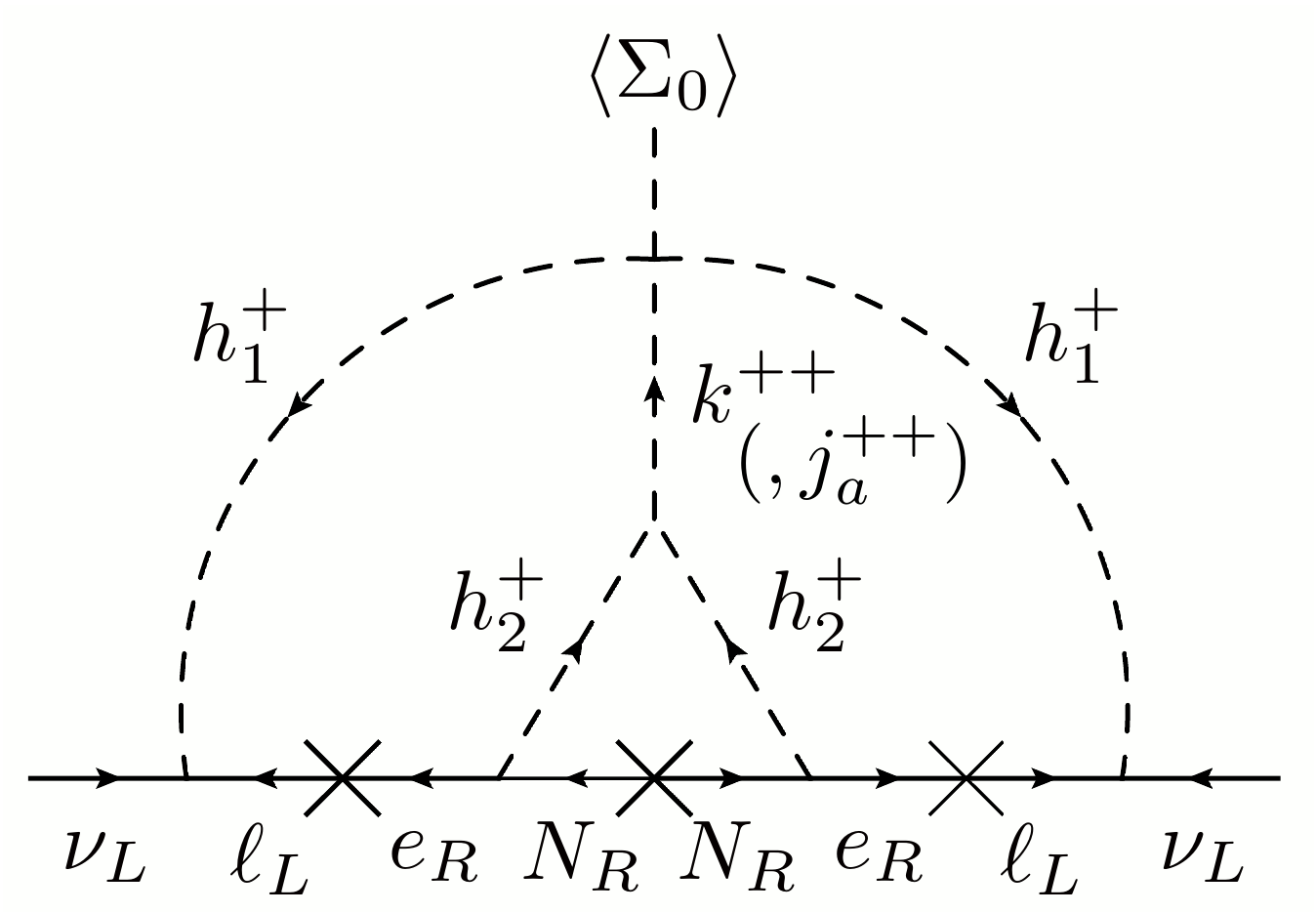}
   \caption{{A schematic description for the radiative generation of neutrino masses.}}
   \label{neutrino-diag}
\end{center}
\end{figure}

We assume that  only the SM-like Higgs $\Phi$ and the additional neutral scalar $\Sigma_0$ have
VEVs, which are symbolized as $\langle\Phi\rangle\equiv v/\sqrt2$ and $\langle\Sigma_0\rangle\equiv v'/\sqrt2$, respectively.
$x\,(\neq0)$ is an arbitrary number of the charge of the hidden $U(1)$ symmetry,
and under the assignments, neutrino mass matrix is generated at the three loop level, where a schematic picture is shown in Fig.~\ref{neutrino-diag}.
A remnant $Z_2$ symmetry remains after the hidden $U(1)$ symmetry breaking and
the particles $N_{R_{1,2,3}}$ and $h^\pm_2$ have negative parities.
Then, when a Majorana neutrino is the lightest among them, it becomes a dark matter~(DM) candidate and the stability is accidentally ensured.

In the original model, {the Lagrangian of Yukawa sector $\mathcal{L}_{Y}$ and scalar potential $\mathcal{V}$}{,} allowed under the gauge and global symmetries{,} are given as
\begin{align}
{-} \mathcal{L}_{Y}
&=
(y_\ell)_{ij} \bar L_{L_i} \Phi e_{R_j}  + \frac12 (y_{L})_{ij} \bar L^c_{L_i} L_{L_j} h^+_1  + (y_{R})_{ij} \bar N_{R_i} e^c_{R_j} h_2^{-}   +  \frac12 (y_N)_{ij} \Sigma_0 \bar N_{R_i}^c N_{R_j} 
+\rm{h.c.} {,} \\ 
\mathcal{V}
&= 
 m_\Phi^2 |\Phi|^2 + m_{\Sigma}^2 |\Sigma_0|^2 + m_{h_1}^2 |h^+_1|^2  + m_{h_2}^2 |h_2^{+}|^2   + m_{k}^2 |k^{++}|^2 
 \notag \\
&+ \Bigl[
 \lambda_{11}  \Sigma_0^* h^-_1 h^-_1 k^{++} +  \mu_{22} h^+_2 h^+_2 k^{--}   + {\rm h.c.}
 \Bigr]
  +\lambda_\Phi |\Phi|^{4} 
  + \lambda_{\Phi\Sigma} |\Phi|^2|\Sigma_0|^2 
 +\lambda_{\Phi h_1}  |\Phi|^2|h^+_1|^2  
 \notag \\
& +\lambda_{\Phi h_2}  |\Phi|^2|h^+_2|^2 +\lambda_{\Phi k}  |\Phi|^2|k^{++}|^2  
  + \lambda_{\Sigma} |\Sigma_0|^{4} +
 \lambda_{\Sigma h_1}  |\Sigma_0|^2|h^+_1|^2  
  +\lambda_{\Sigma h_2}  |\Sigma_0|^2|h^+_2|^2 
   \notag \\&
 +\lambda_{\Sigma k}  |\Sigma_0|^2|k^{++}|^2  
  + \lambda_{h_1} |h_1^{+}|^{4}
  {+} \lambda_{h_1 h_2}  |h_1^{+}|^2|h^+_2|^2 +\lambda_{h_1 k}  |h_1^{+}|^2|k^{++}|^2  
 \notag \\
&+
 \lambda_{h_2} |h_2^{+}|^{4} + \lambda_{h_2 k}  |h_2|^2|k^{++}|^2  
+ \lambda_{k} |k^{++}|^{4} 
,
\label{HP}
\end{align}
where the indices $i,j$ indicate matter generations and the superscript $``c"$ means charge conjugation (with the $SU(2)_L$ rotation by $i\sigma_2$ for $SU(2)_L$ doublets).
We assume that $y_N$ is diagonal, where the right-handed neutrino masses are calculated as $M_{N_i} = \frac{v'}{\sqrt{2}} (y_N)_{ii}$ with the assumed ordering $M_{N_1} (=\text{DM mass}) < M_{N_2} < M_{N_3}$.
The neutral scalar fields are shown in the Unitary gauge as 
\begin{align}
&\Phi =\left[
\begin{array}{c}
0\\
\frac{v+\phi}{\sqrt2}
\end{array}\right],\quad 
\ 
\Sigma_0=\frac{v'+\sigma}{\sqrt{2}}e^{iG/v'},
\label{component}
\end{align}
with $v \simeq 246\,\text{GeV}$ and an associated Nambu-Goldstone~(NG) boson $G$ via the global $U(1)$ breaking due to the occurrence of nonzero $v'$.
Requiring the tadpole conditions, {$\partial\mathcal{V}/\partial\phi|_{\phi = v} = \partial\mathcal{V}/\partial\sigma|_{\sigma = v'} = 0$},
{the} {resultant} mass matrix squared of the {$CP$} even components $(\phi,\sigma)$ 
 is given by
\begin{equation}
m^{2} (\phi,\sigma) = \left[%
\begin{array}{cc}
  2\lambda_\Phi v^2 & \lambda_{\Phi \Sigma}vv' \\
  \lambda_{\Phi \Sigma}vv' & 2\lambda_{\Sigma}v'^2 \\
\end{array}%
\right] = \left[\begin{array}{cc} \cos\alpha & \sin\alpha \\ -\sin\alpha & \cos\alpha \end{array}\right]
\left[\begin{array}{cc} m^2_{h} & 0 \\ 0 & m^2_{H}  \end{array}\right]
\left[\begin{array}{cc} \cos\alpha & -\sin\alpha \\ \sin\alpha &
      \cos\alpha \end{array}\right],
      \label{eq:CP-even_matrix}
\end{equation}
where $h$ is the SM-like Higgs {($m_h = 125\,\text{GeV}$)} and $H$ is an additional {$CP$ even} Higgs mass eigenstate.
The mixing angle $\alpha$ is {determined as} 
\al{
\sin 2\alpha=\frac{2\lambda_{\Phi \Sigma} v v'}{{m^2_H-m_h^2}}.
\label{eq:CP-even_mixing}
}
The neutral bosons $\phi$ and $\sigma$ are {represented} in terms of the mass eigenstates $h$ and $H$ as
\begin{eqnarray}
\phi = h\cos\alpha + H\sin\alpha,
\quad
\sigma =- h\sin\alpha + H\cos\alpha.
\label{eq:mass_weak}
\end{eqnarray}
The two $CP$ even scalars {$h$ and $H$} could work as DM-portal scalars and participate in the DM pair annihilation.
The mass {eigenvalues} for the {singly charged} bosons $h_1^\pm$, $h_2^\pm$ and the {doubly charged boson} $k^{\pm\pm}$ are given as
\begin{align}
&m^{2}_{h^{\pm}_1} = m_{h_1}^{2}  + \frac12 (\lambda_{\Phi h_1} v^{2}+\lambda_{\Sigma h_1} v'^{2}), \quad 
m^{2}_{h^{\pm}_2} = m_{h_2}^{2}  + \frac12 (\lambda_{\Phi h_2} v^{2}+\lambda_{\Sigma h_2} v'^{2}), \notag \\ 
&m^{2}_{k^{\pm\pm}} = m_{k}^{2}  + \frac12 (\lambda_{\Phi k} v^{2}+\lambda_{\Sigma k} v'^{2}).
	\label{eq:mass_eigenvalues}
\end{align}

This model can explain the smallness of the observed neutrino masses and the presence of DM without severe parameter tuning.
A summary of the features in the model is given in {Appendix~\ref{sec:app:review}}.

Here we introduce a real singlet scalar $S$ in the model  and assume that it couples with the doubly charged scalar(s).  {Due to the contributions of the charged particles in the loop,} a large branching ratio $\B(S \to \gamma\gamma)$ {is achievable without assuming tree level interactions~\cite{Fichet:2015vvy,Csaki:2015vek}.} 
{When $\B(S \to \gamma\gamma)$ is sizable,  the production cross section of the resonance particle, $\sigma(pp\to S+X)$, becomes large through photon fusion processes thus we do not have to rely on gluon fusion processes, which often requests additional colored particles that brings in dangerous hadronic activities.} {Thus we may explain the 750 GeV excess as pointed out in~\cite{Fichet:2015vvy,Csaki:2015vek}.}

\subsection{Extension with a scalar $S$  for the 750 GeV resonance}

In the following part, we consider an extension of the original model with the {new interactions as}
\al{
\Delta \mathcal{V} &=
{\hat{\mu}_{Sk} S |k^{++}|^2 + {\hat{\lambda}_{S k}} S^2 |k^{++}|^2 + \V(S)} \notag \\
&+ \sum_{a=1}^{N_j} 
\left\{
  {\hat{m}_{j_a^{\pm\pm}}^2} |j^{++}_a|^2
{+ \hat{\mu}_{S j_a} S |j^{++}_a|^2 + {\hat{\lambda}_{S j_a}} S^2 |j^{++}_a|^2}
+ \left[ {\lambda_{11}^{(a)}} \Sigma_0^\ast h^-_1 h^-_1 j_a^{++} + {\mu_{22}^{(a)}} h^+_2 h^+_2 j_a^{--} + \text{h.c.} \right]
\right\},
	\label{eq:additional_terms}
}
where $S$ is a real scalar and $j^{\pm\pm}_a\,(a=1,2,\cdots,N_j)$ are {additional} $SU(2)_L$ singlet doubly charged scalars with hyper charge $+2$ and a global $U(1)$ charge $+2x$.
{$\V(S)$ represents the potential of the singlet scalar $S$.
Here, we assume that $S$ has a VEV, and $S$ should be replaced as $S \to \langle S \rangle + S$.
After the replacement, we pick up the relevant terms for our analysis and summarize,}
\al{
\Delta \mathcal{V}_{\text{eff}} &=
{\mu_{Sk} S |k^{++}|^2 + \frac{1}{2} m_S^2 S^2} \notag \\
&+ \sum_{a=1}^{N_j} 
\left\{
  m_{j_a^{\pm\pm}}^2 |j^{++}_a|^2
{+ \mu_{S j_a} S |j^{++}_a|^2}
+ \left[ {\lambda_{11}^{(a)}} \Sigma_0^\ast h^-_1 h^-_1 j_a^{++} + {\mu_{22}^{(a)}} h^+_2 h^+_2 j_a^{--} + \text{h.c.} \right]
\right\},
	\label{eq:additional_terms_effective}
}
with
\al{
{
m_{j_a^{\pm\pm}}^2 \equiv \hat{m}_{j_a^{\pm\pm}}^2 + {\hat{\mu}_{S j_a}} \langle S \rangle
+ {\hat{\lambda}_{S j_a}} \langle S \rangle^2,\quad
\mu_{Sk} \equiv \hat{\mu}_{Sk} + 2 {\hat{\lambda}_{S k}} \langle S \rangle, \quad
\mu_{Sj_a} \equiv \hat{\mu}_{Sj_a} + 2 {\hat{\lambda}_{S j_a}} \langle S \rangle.}
	\label{eq:effective_scalar_trilinear}
}
The squared physical masses of $S$ and $j^{\pm\pm}_a$ are $m_S^2$ and $m_{j_a^{\pm\pm}}^2$, respectively and we set $m_S$ as $750\,\text{GeV}$ for our explanation of the $750\,\text{GeV}$ excess.\footnote{
In a later stage of Sec.~\ref{sec:withmixing}, we have discussions on the situation when $S$ and $\Phi$ are mixed.}
$j_a^{\pm\pm}$ has the same charges as $k^{\pm\pm}$ and then can contribute to the three-loop induced neutrino masses shown in Fig.~\ref{neutrino-diag}.\footnote{
{In general, mixing between $k^{\pm\pm}$ and $j_{a}^{\pm\pm}$ could be allowed but the induced value via renormalization group running at the scale of our interest is expected to be small with heavy masses of $h_1^\pm$ and $h_2^\pm$, thus is neglected.}
}
The trilinear terms {in the square brackets} are required for evading the stability of $j_a^{\pm\pm}$.
{We also ignore the possible terms as $|j_a^{++}|^2 |\Phi|^2$, $|j_a^{++}|^2 |\Sigma_0|^2$ and $S |\Phi|^2$, $S |\Sigma_0|^2$ in Eq.~(\ref{eq:additional_terms}) in {our} analysis below.
This is justified as a large VEV of $S$ generates large effective trilinear couplings $\mu_{S k}$ and $\mu_{S j_a}$ through the original terms $S^2 |k^{++}|$ and $S^2 |j^{++}_a|$, respectively{,} even when the dimensionless coefficients {$\hat{\lambda}_{S k}$} and {$\hat{\lambda}_{S j_a}$} are not large.
}

\section{Analysis
\label{sec:analysis}}

\subsection{Formulation of ${p(\gamma) p(\gamma)} \to S + X \to \gamma \gamma + X$}

Additional interactions in Eq.~(\ref{eq:additional_terms_effective}) provide possible decay channels of $S$ to $\gamma\gamma$, ${Z\gamma}$, $ZZ$ and $k^{++} k^{--}$ or $j_a^{++} j_a^{--}$ up to the one-loop level. 
We assume that {$m_{k^{\pm\pm}}$ and $m_{j_a^{\pm\pm}}$ are} greater than $m_S/2\,(=375\,\text{GeV})$, where the last two decay channels {at the tree level} are closed kinematically.
{Here, we show the case when $S$ is a mass eigenstate and there is no mixing through mass terms with other scalars.}
In the present case that {no tree-level decay branch is open and} only $SU(2)_L$ singlet charged scalars describe the {loop-induced} partial widths, the relative strengths among $\Gamma_{S \to \gamma\gamma}$, $\Gamma_{S \to {Z\gamma}}$, $\Gamma_{S \to ZZ}$ and $\Gamma_{S \to W^+ W^-}$ are governed by quantum numbers at the one-loop level\footnote{{The branching fractions are easily understood in an effective theory with the standard model gauge symmetries. See e.g. \cite{Kim:2015vba} with $s_2=0$ in the paper.}} 
 as
\al{
\Gamma_{S \to \gamma\gamma} :
\Gamma_{S \to {Z \gamma}} :
\Gamma_{S \to ZZ} :
\Gamma_{S \to W^+ W^-}
\approx
1 :
2 \left(\frac{s_W^2}{c_W^2}\right) :
\left(\frac{s_W^4}{c_W^4}\right) :
0.
	\label{eq:oneloopGamma_ratios}
}
In the following, we calculate $\Gamma_{S \to ZZ}$ in a simplified way of
\al{
\Gamma_{S \to ZZ} \approx s_W^2/(2c_W^2) \Gamma_{S \to {Z \gamma}} \simeq 0.15 \, \Gamma_{S \to {Z \gamma}}.
	\label{eq:StoVV_1}
}

Here, we represent a major part of partial decay widths of $S$ with our notation for loop functions with the help of Refs.~\cite{Ellis:1975ap,Shifman:1979eb,Djouadi:2005gi,Carena:2012xa,Chen:2013vi}.
In the following part, for simplicity, we set all the masses of the doubly charged scalars $m_{j_a^{\pm\pm}}$ as the same as $m_{k^{\pm\pm}}$, while we ignore the contributions from the two singly charged scalars $h_{1,2}^\pm$ since they should be heavy as at least around $3\,\text{TeV}$ and decoupled as mentioned in {Appendix~\ref{sec:app:review}}.
{The concrete form of $\Gamma_{S \to \gamma\gamma}$ and $\Gamma_{S \to {Z \gamma}}$ are given as}
\begin{align}
\Gamma_{S \to \gamma\gamma}
&=
\frac{\alpha_{\text{EM}}^2 {m_S^3}}{{256}\pi^3 v^2}
\left|
\frac{1}{2} \frac{v \mu}{m_{k^{\pm\pm}}^2} Q_k^2 A_0^{\gamma\gamma}(\tau_k)
\right|^{2},
	\label{eq:StoVV_2} \\
\Gamma_{S \to Z\gamma}
&=
\frac{{\alpha_{\text{EM}}^2} {m_{S}^3}}{512\pi^3} \left( 1 - \frac{m_Z^2}{{m_S^2}} \right)^3
\left|
- \frac{\mu}{m_{k^{\pm\pm}}^2} \left( 2 Q_k g_{Zkk} \right) A_0^{Z\gamma}(\tau_k,\lambda_k)
\right|^2,
	\label{eq:StoVV_3}
\end{align}
with
\begin{align}
\mu =\sum_a \mu_a \equiv
	\left[ {\mu_{S k}} + \sum_{a=1}^{N_j} \mu_{S j_a} \right], \quad
g_{Zkk} = - Q_k \left( \frac{s_W}{c_W} \right), \quad
\tau_k = \frac{4 m_{k^{\pm\pm}}^2}{m_S^2}, \quad
\lambda_k = \frac{4 m_{k^{\pm\pm}}^2}{m_Z^2}.
	\label{effective_couplings_oneloopHiggs}
\end{align}
$Q_k\,(=2)$ is the electric {charge} of the doubly charged scalars in unit of the positron's one.
$c_W$ and $s_W$ are the cosine and the sine of the Weinberg angle $\theta_W$, respectively.
{$\alpha_{\text{EM}}$ is the electromagnetic fine structure constant.}
In the following calculation, we use $s_W^2 = 0.23120$ and $\alpha_{\text{EM}} = 1/127.916$.
The loop factors take the following forms,
\begin{align}
A_{0}^{\gamma\gamma}(x)
&=
-x^2 \left[ x^{-1} - f(x^{-1}) \right], \notag \\
A_{0}^{Z\gamma}(x,y) 
&=
\frac{xy}{2(x-y)} + \frac{x^2y^2}{2(x-y)^2} \left[ f(x^{-1}) - f(y^{-1}) \right] + \frac{x^2y}{(x-y)^2} \left[ g(x^{-1}) - g(y^{-1}) \right],
	\label{Azerofunctions}
\end{align}
The two functions $f(z)$ and $g(z)$ ($z \equiv x^{-1}\ \text{or}\ y^{-1}$) are formulated as
\begin{align}
f(z) &=
\arcsin^2{\sqrt{z}} \quad \text{for } z \leq 1,
	\label{f(z)form} \\
g(z) &=
\sqrt{z^{-1} - 1} \arcsin{\sqrt{z}} \quad \text{for } z \leq 1,
	\label{g(z)form}
\end{align}
where the situation $m_{S} \leq 2 m_{k^{\pm\pm}},\, m_{Z} \leq 2 m_{k^{\pm\pm}}$ corresponds to $z \leq 1$.
For simplicity, we assume the relation
\al{
\mu_{Sk} = \mu_{Sj_a},
}
for all $a$.

For the production of $S$ corresponding to the $750\,\text{GeV}$ resonance, we consider the photon fusion process, as firstly discussed in the context of the $750\,\text{GeV}$ excess in Refs.~\cite{Fichet:2015vvy,Csaki:2015vek}.
We take the photon parton distribution function (PDF) from Ref.~\cite{Harland-Lang:2016qjy}, which adopted the methods in {Ref.}~\cite{Martin:2014nqa}.\footnote{
See also~\cite{Franceschini:2015kwy,Anchordoqui:2015jxc,
Danielsson:2016nyy,Nomura:2016fzs,Csaki:2016raa,Ito:2016zkz,D'Eramo:2016mgv,Sahin:2016lda,Fichet:2016pvq,Harland-Lang:2016apc,Franzosi:2016wtl,Abel:2016pyc,Nomura:2016rjf,Ben-Dayan:2016gxw,Martin:2016byf,Barrie:2016ntq,Ito:2016qsm,Gross:2016ioi,Baek:2016uqf,Ko:2016sxg,Molinaro:2016oix,Panico:2016ary,Bharucha:2016jyr,Ababekri:2016kkj,Anchordoqui:2016rve,Howe:2016mfq,Frandsen:2016bke} for related issues.
}
The inclusive production cross section of a scalar (or pseudoscalar) resonance $R$ is generally formulated as
\al{
\frac{d \sigma^{\text{inc}}(p(\gamma) p(\gamma) \to R + X) }{d M_{R}^{2} \, d y_{R}}
=
\frac{d \mathcal{L}^{\text{inc}}}{d M_{R}^{2} \, d y_{R}}
\hat{\sigma} (\gamma \gamma \to R),
}
where $M_{R}$ and $y_{R}$ are the mass and the rapidity of the resonance $R$, and $\hat{\sigma} (\gamma \gamma \to R)$ shows the parton-level cross section for the process $\gamma \gamma \to R$.
The inclusive luminosity function {can be conveniently written} in terms of the photon PDF as
\al{
\frac{d \mathcal{L}^{\text{inc}}_{\gamma\gamma}}{d M_{R}^{2} \, d y_{R}}
=
\frac{1}{s} \gamma(x_{1}, \mu) \, \gamma(x_{2}, \mu),
}
where $x_{1,2} = \frac{M_{R}}{\sqrt{s}} e^{\pm y_{R}}$ represent the momentum fractions of the photons inside the protons and $\sqrt{s}$ means the total energy. {The value of $\gamma(x, \mu)$ can be evaluated by taking the Dokshitzer-Gribov-Lipatov-Altarelli-Parisi~(DGLAP) evolution from the starting scale $\mu_{0} \, (=1\,\text{GeV})$ to $\mu$ after an estimation of coherent and incoherent components of the initial form of $\gamma(x, \mu=\mu_0)$ at $\mu = \mu_{0}$ (See \cite{Harland-Lang:2016qjy} for details.).}

By adopting the narrow width approximation, {which is fine in our case}, the parton-level cross section of the particle $S$ of mass $m_{S}$ and rapidity $y_{S}$ is given as
\al{
\hat{\sigma} (\gamma \gamma \to S) &= \frac{8\pi^{2} \Gamma(S \to \gamma\gamma)}{m_{S}} \delta(M_{R}^{2} - m_{S}^{2})
\notag \\
&= \frac{8\pi^{2} \Gamma_{\text{tot}}(S)}{m_{S}} \B(S \to \gamma\gamma) \delta(M_{R}^{2} - m_{S}^{2}).
}
{The inclusive differential cross section is obtained in a factorized form:}
\al{
\frac{d \sigma^{\text{inc}}(p(\gamma) p(\gamma) \to S + X) }{d y_{S}}
=
\left.
\frac{8\pi^{2} \Gamma(S \to \gamma\gamma)}{m_{S}} \times
\frac{d \mathcal{L}^{\text{inc}}_{\gamma\gamma}}{d M_{R}^{2} \, d y_{S}} \right|_{M_{R} = m_{S}}.
}

{Now taking the values for $\gamma(x,\mu)$ in Ref.~\cite{Harland-Lang:2016qjy}, we obtain a convenient form of cross section}
\al{
\sigma^{\text{inc}}(p(\gamma) p(\gamma) \to S + X)
=
91\,\text{fb} \left( \frac{\Gamma_{\text{tot}}(S)}{1\,\text{GeV}} \right) \B(S \to \gamma\gamma),
\label{eq:photoproduction_total}
}
or
\al{
\sigma^{\text{inc}}(p(\gamma) p(\gamma) \to S + X \to \gamma \gamma + X)
=
91\,\text{fb} \left( \frac{\Gamma_{\text{tot}}(S)}{1\,\text{GeV}} \right) \B^{2}(S \to \gamma\gamma),
\label{eq:photoproduction_togammagamma}
}
for evaluating production cross sections at $\sqrt{s} = 13\,\text{TeV}$.
The reference magnitude of the cross section, {91\,\text{fb},} is much greater than that in~\cite{Csaki:2015vek} obtained under the narrow width approximation and effective photon approximation~\cite{vonWeizsacker:1934nji,Williams:1934ad}, $1.6-3.6 \,\text{fb}$ (depending on the minimum impact parameter for elastic scattering), while it is smaller than that in~\cite{Csaki:2016raa} through a similar calculation with in~\cite{Harland-Lang:2016qjy}, $240\,\text{fb}$.
We also find at $M_{R} = 750\,\text{GeV}$ in Ref.~\cite{Harland-Lang:2016qjy}
\al{
\frac{\mathcal{L}^{\text{inc}}_{\gamma\gamma}(\sqrt{s} = 13\,\text{TeV})}{\mathcal{L}^{\text{inc}}_{\gamma\gamma}(\sqrt{s} = 8\,\text{TeV})} \approx 2.9.
\label{eq:luminosity_ratio}
}
Having the above relations in Eqs.~(\ref{eq:photoproduction_total})-(\ref{eq:luminosity_ratio}), it is straightforward to evaluate the inclusive production cross section at $\sqrt{s} = 8\,\text{TeV}$.
We note that the resultant value is greater than the value ($\approx 2$) cited in {Ref.}~\cite{Csaki:2016raa}.

\subsection{Results}

\subsubsection{Case 1: without mass mixing}
\label{sec:withoutmixing}

\begin{table}[t]
\begin{tabular}{|c|c|c|c|}\hline  
Final state & {Upper bound} (in fb, $95\%$ C.L.) & Category & Ref. \\ \hline
$\gamma\gamma$ & $2.4$/$2.4$ & 8TeV-ATLAS/CMS & \cite{Aad:2015mna,CMS-PAS-HIG-14-006} \\
               & $13$/$13$ & 13TeV-ATLAS/CMS & \cite{ATLAS-CONF-2016-018,CMS-PAS-EXO-16-018} \\ \hline
$Z\gamma$      & $4.0$/$27$ & 8TeV/13TeV-ATLAS & \cite{Aad:2014fha,ATLAS-CONF-2016-010} \\ \hline
$ZZ$           & $12$/$99$ & 8TeV/13TeV-ATLAS & \cite{Aad:2015kna,ATLAS-CONF-2016-016} \\ \hline
$WW$           & $35$   & 8TeV-ATLAS & \cite{Aad:2015agg} \\ \hline
$hh$           & $40$   & 8TeV-ATLAS & \cite{ATLAS-CONF-2014-005} \\ \hline
\end{tabular}
\centering
\caption{{$95\%$ C.L. upper bounds} on decay channels of a $750\,\text{GeV}$ scalar resonance.
}
\label{tab:LHCconstraints_on_S}
\end{table}

{In this part, we discuss the case that the field $S$ is a mass eigenstate, where no mixing effect is present through mass terms with other scalars.}
Under our assumptions, the relevant parameters are {$(m_{k^{\pm\pm}}, \mu_{S k}, N_j)$: the universal physical mass of the doubly charged scalars (assuming $m_{k^{\pm\pm}} = m_{j_{a}^{\pm\pm}}$ for all $a$), the universal effective scalar trilinear coupling (assuming $\mu_{S k} = \mu_{S j_{a}}$ for all $a$), and the number of the additional doubly charged singlet scalars.}
We observe the {unique} relation among the branching ratios of $S$ {irrespective of $m_{k^{\pm\pm}}$ and $\mu_{S k}$}, {which is suggested by} Eq.~(\ref{eq:oneloopGamma_ratios}), as
\al{
\B(S \to \gamma\gamma) \simeq {0.591},\quad
\B(S \to \gamma Z) \simeq {0.355},\quad
\B(S \to Z Z) \simeq {0.0535}.
	\label{eq:S_universal_B}
}

In Ref.~\cite{Kamenik:2016tuv}, reasonable target values for the cross section of $\sigma_{\gamma\gamma} \equiv \sigma(p p \to {S + X} \to {\gamma\gamma +X})$ at the $\sqrt{s} = 13\,\text{TeV}$ LHC were discussed as functions of the variable $R_{13/8}$, which is defied as
\al{
R_{13/8} \equiv \frac{\sigma(pp \to S)|_{\sqrt{s} = 13\,\text{TeV}}}{\sigma(pp \to S)|_{\sqrt{s} = 8\,\text{TeV}}},
}
where the published data after Moriond 2016 are included, and the four categories discriminated by the two features (spin-$0$ or spin-$2$; narrow width [$\Gamma_{S}/m_{S} \to 0$] or wide width [$\Gamma_{S}/m_{S} = 6\%$]) are individually investigated.
As pointed out in Eq.~(\ref{eq:S_universal_B}), the value of $\B(S \to \gamma\gamma)$ is {uniquely} fixed as {$\simeq 60\%$} and $S$ is produced only through the photon fusion in the present case.
As shown in Eq.~(\ref{eq:luminosity_ratio}) in our estimation of the photo-production, $R_{13/8}$ corresponds to $2.9$,
where the best fit values of $\sigma_{\gamma\gamma}$ at ${\sqrt{s} = 13\,\text{TeV}}$ are extracted from~\cite{Kamenik:2016tuv} as
\al{
2.0 \pm 0.5 \, \text{fb} \quad (\text{for } \Gamma_{S}/m_{S} \to 0),\qquad
4.25 \pm 1.0 \, \text{fb} \quad (\text{for } \Gamma_{S}/m_{S} = 6\%).
	\label{eq:bestfit_range_1}
}
The theoretical error in the present formulation of the photo-production was evaluated as $\pm 15-20\%$ in~\cite{Harland-Lang:2016qjy}.
Then, we decide to focus on the $2\sigma$ favored regions with including the error ($20\%$, fixed) also, concretely speaking,
\al{
[0.8, 3.6] \, \text{fb} \quad (\text{for } \Gamma_{S}/m_{S} \to 0),\qquad
[1.8, 7.5] \, \text{fb} \quad (\text{for } \Gamma_{S}/m_{S} = 6\%).
	\label{eq:bestfit_range_2}
}
Here, the $95\%$ CL upper bound on $\sigma_{\gamma\gamma}$ at ${\sqrt{s} = 8\,\text{TeV}}$ is {$\lesssim 2.4\,\text{fb}$}~\cite{Aad:2015mna,CMS-PAS-HIG-14-006} and the favored regions are still consistent with the 8\,TeV result (or just on the edge).
{It is found that} the bounds on the $Z\gamma$, $ZZ$ final states are weaker than that of $\gamma\gamma$.
Relevant information is summarized in Table~\ref{tab:LHCconstraints_on_S}.

\begin{figure}[H]
\centering
\includegraphics[width=0.32\columnwidth]{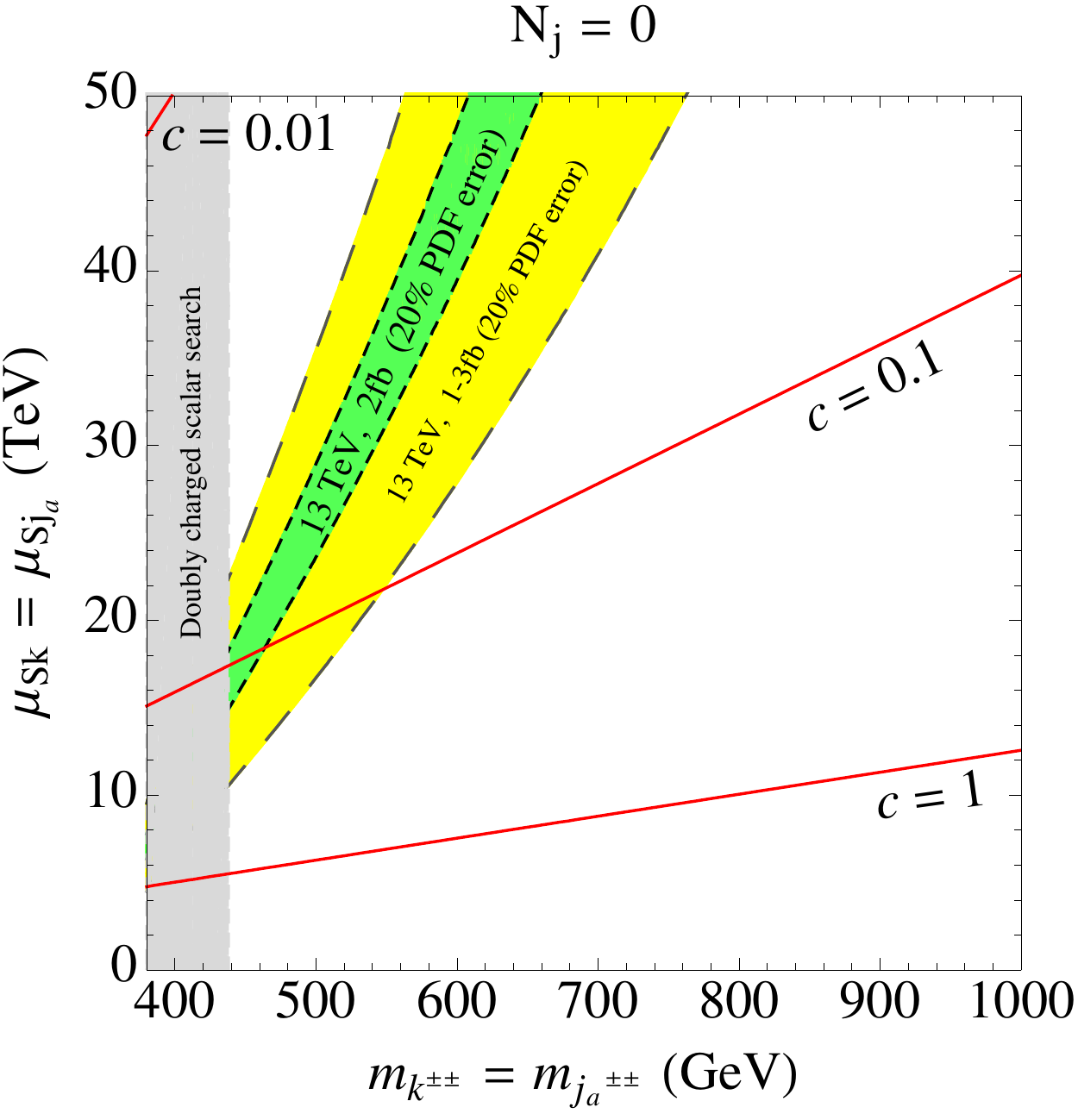}\ 
\includegraphics[width=0.32\columnwidth]{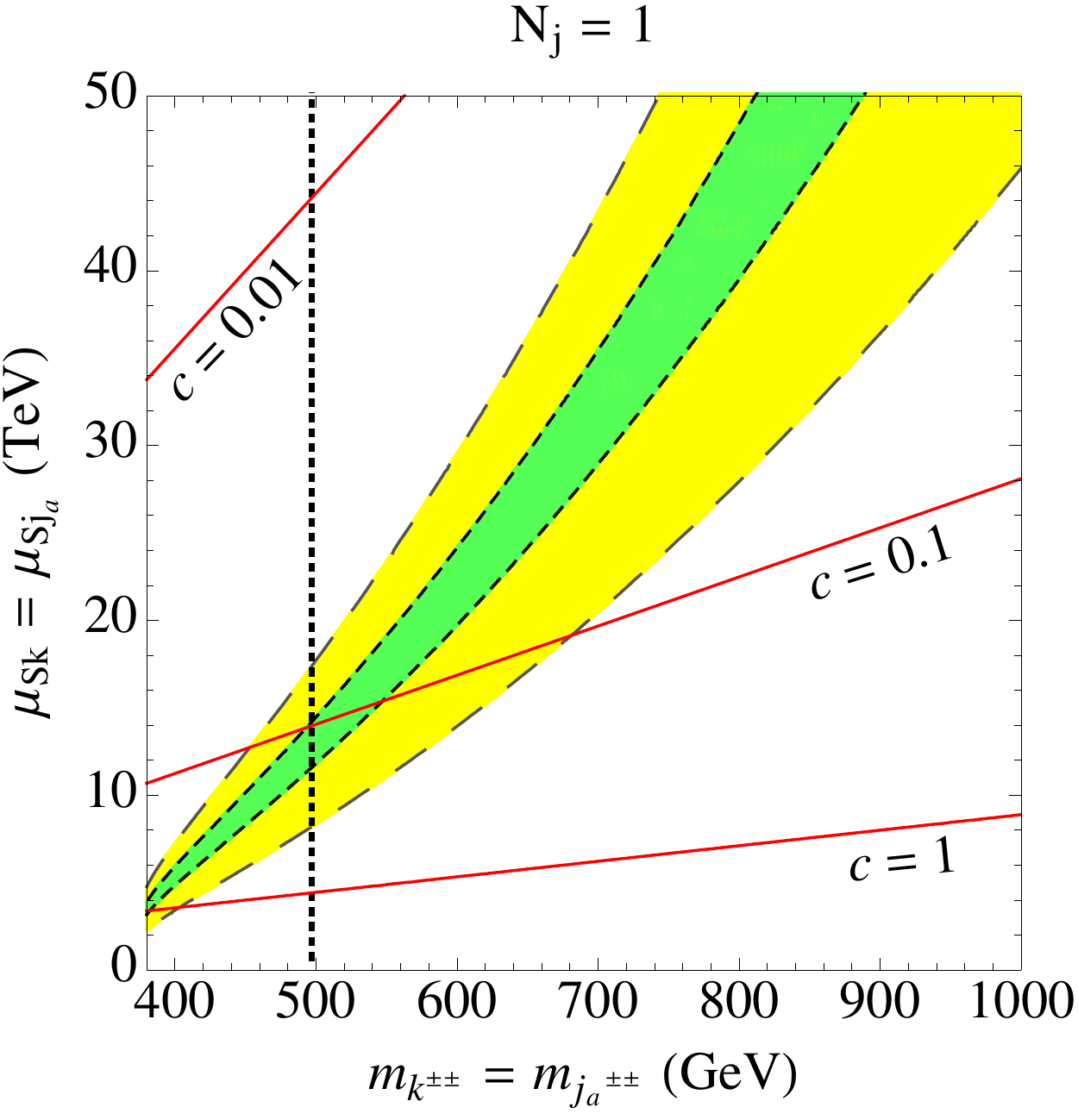}\
\includegraphics[width=0.32\columnwidth]{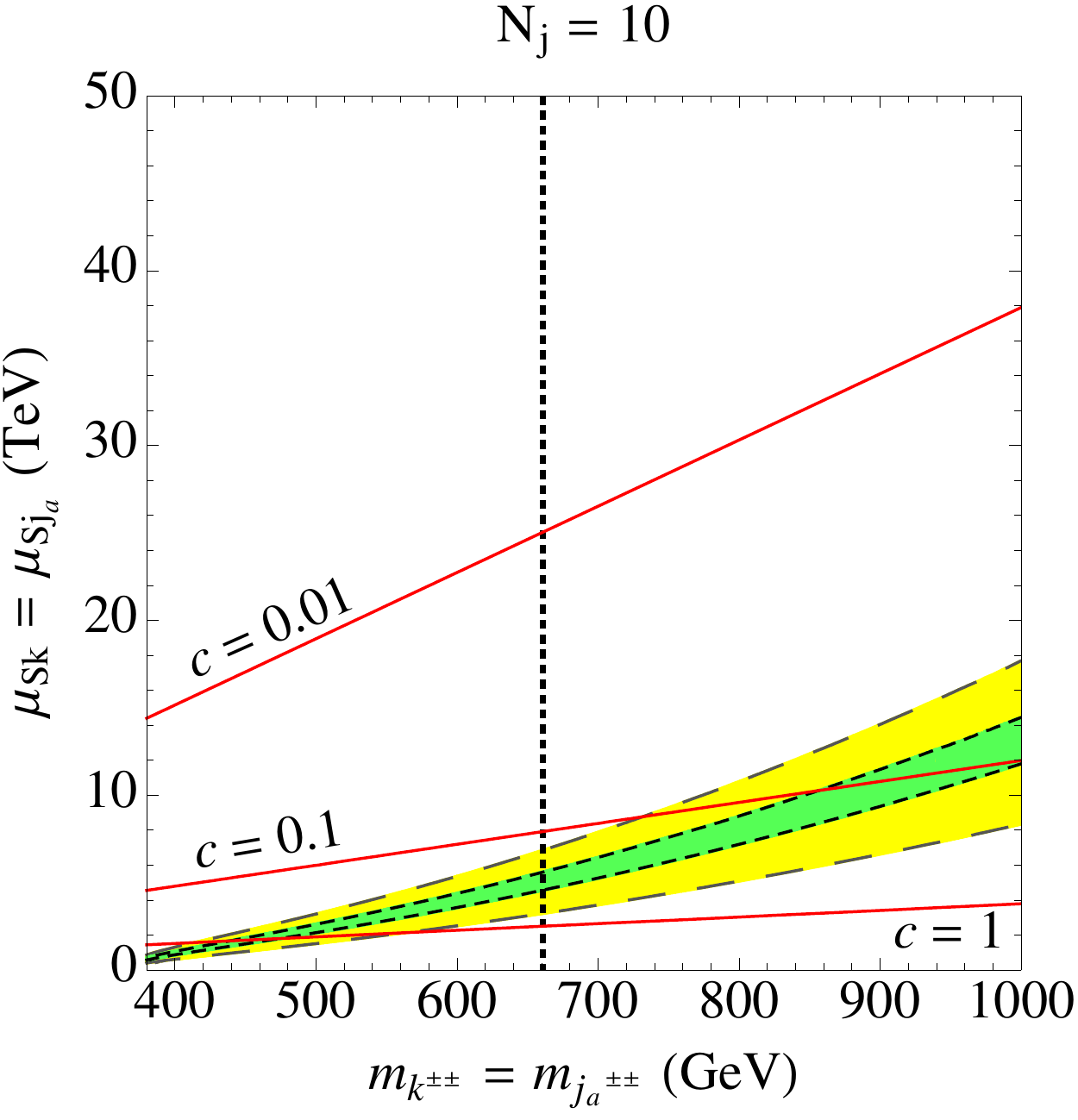} \\[6pt]
\includegraphics[width=0.32\columnwidth]{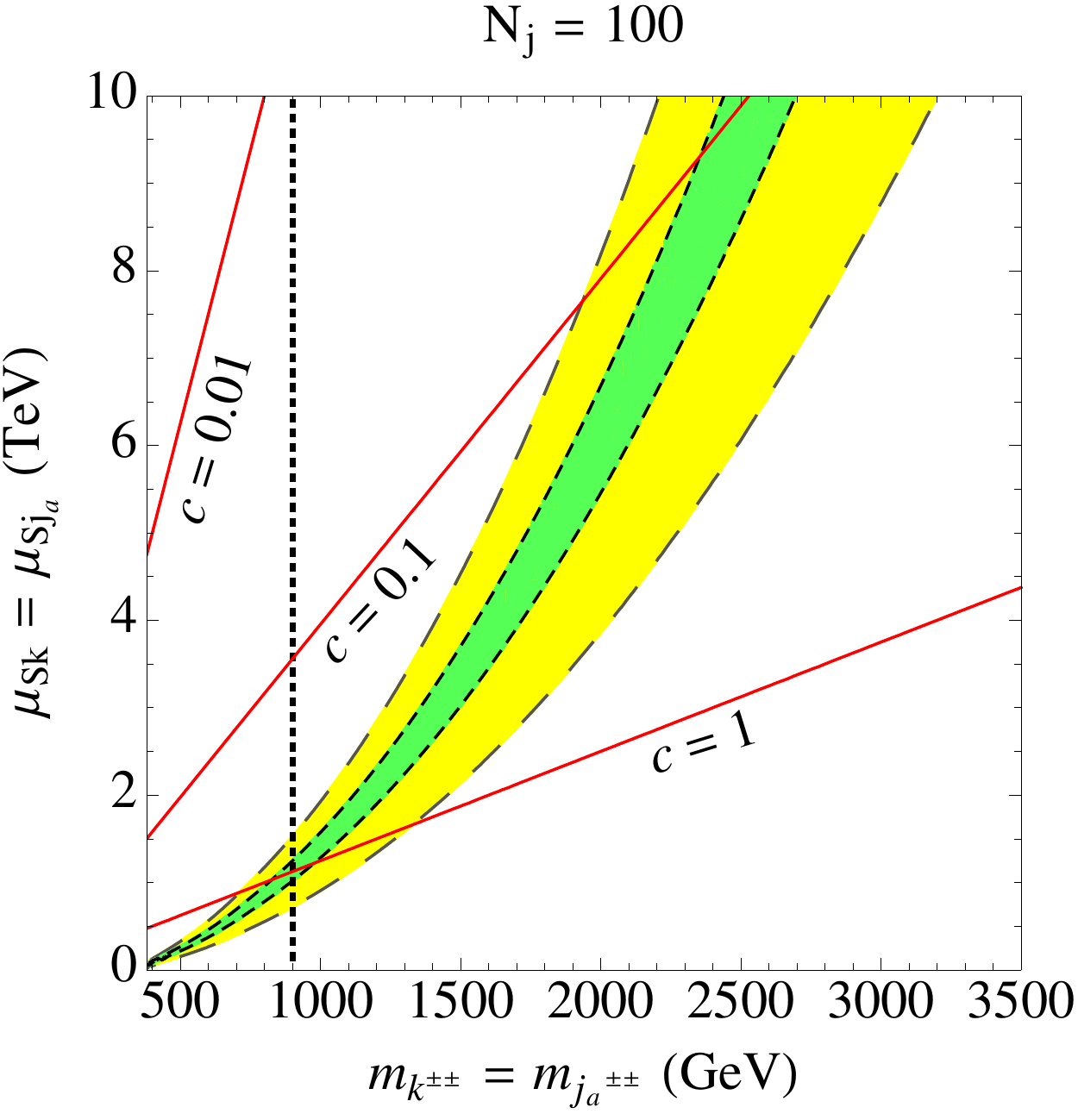}\ 
\includegraphics[width=0.32\columnwidth]{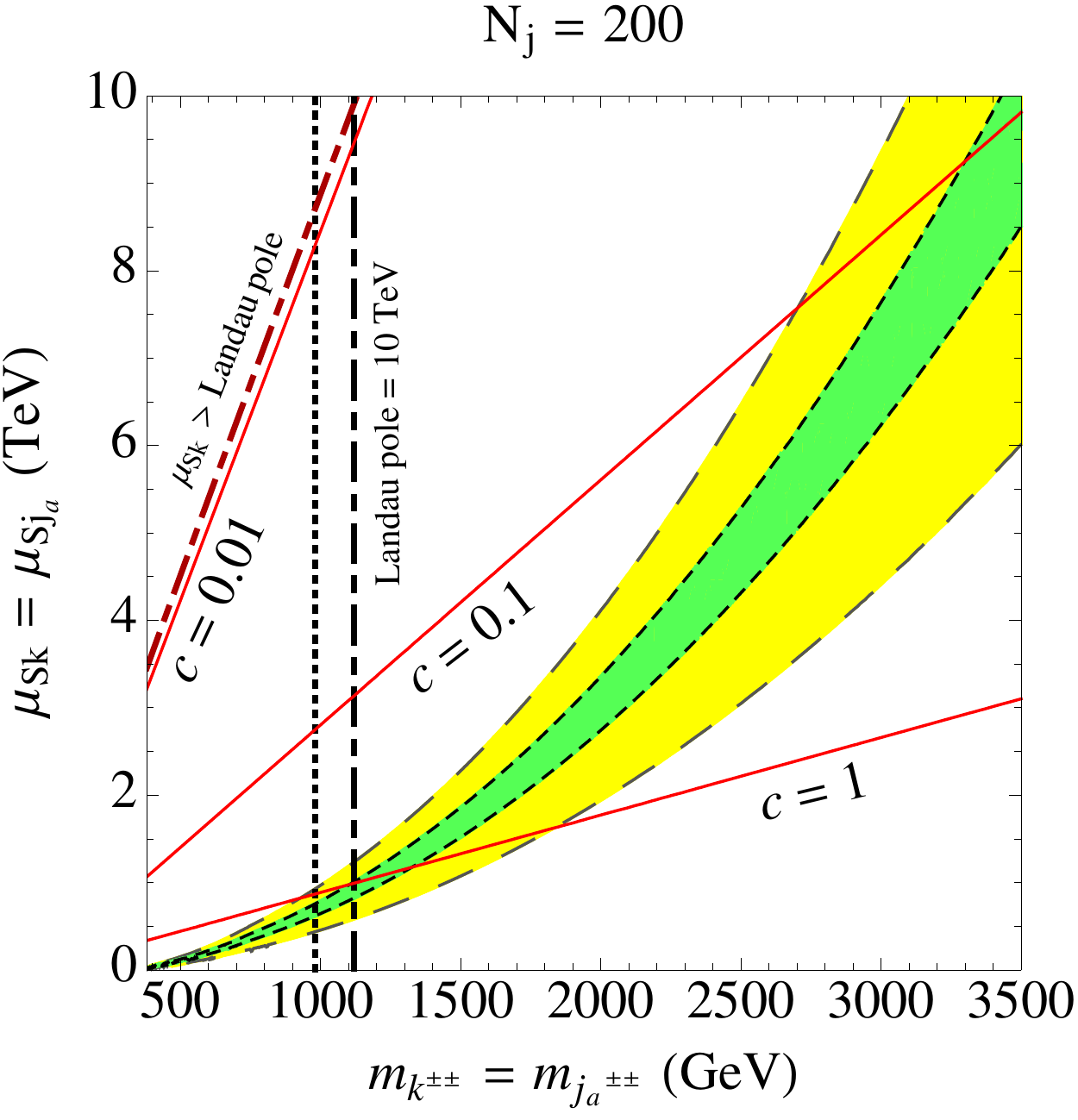}\
\includegraphics[width=0.32\columnwidth]{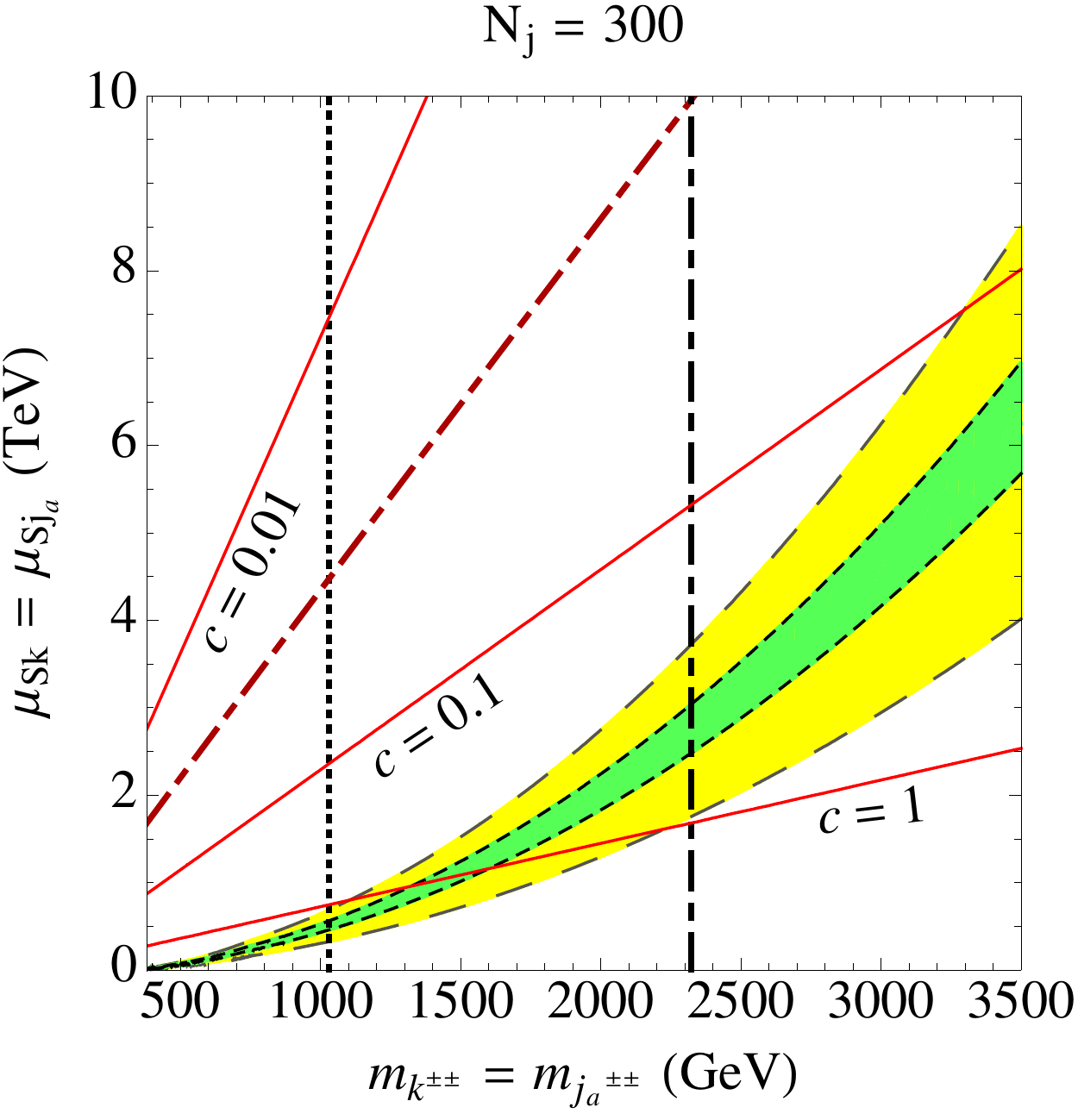} 
\caption{
Six cases with different numbers of doubly charged scalars are considered with $N_j=0,1,10,100,200$ and $300$.
{Inside the green regions, the best fit value of the production cross section is realized with taking account of $\pm 20\%$ theoretical error discussed in~\cite{Harland-Lang:2016qjy}.
The yellow regions indicate the areas where we obtain the $2\sigma$-favored values in the production cross section of $p(\gamma)p(\gamma) \to S + X \to \gamma\gamma +X$, where we take account of both of the theoretical ($\pm 20\%$) and experimental (shown in Eq.~(\ref{eq:bestfit_range_1})) errors.}
Cross section evaluations are owing to Eq.~(\ref{eq:photoproduction_togammagamma}).
The gray shaded region $m_{k^{\pm\pm}} \leq 438\,\text{GeV}$ in $N_j=0$ shows the excluded parts in $95\%$ {C.L.} {via the ATLAS 8\,TeV search for doubly charged particles} with the assumption of $\B(k^{\pm\pm} \to \mu^\pm \mu^\pm) = 100\%$~\cite{ATLAS:2014kca}.
The vertical black {dotted} lines represent corresponding bounds on the universal physical mass $m_{k^{\pm\pm}}$ when we assume $\B(j^{\pm\pm}_a \to \mu^\pm \mu^\pm) = 100\%$ for all $j^{\pm\pm}_a$.
Two types of constraints with respect to ``Landau pole'' of $g_{Y}$ (defined as $g_{Y}(\mu) = 4\pi$) are meaningful when $N_{j}$ is large ($N_{j} = 200,\,300$).
{The red lines indicate three reference boundaries of the correction factor $c \, \delta = 1$ with $c = 1,\,0.1,\,0.01$ to the trilinear couplings $\mu_{S k} \, (= \mu_{S j_{a}})$ defined in Eq.~(\ref{eq:trilinear_correction}).
For each choice of $c$, the region below the corresponding boundary is favored from a viewpoint of perturbativity.}
}
   \label{fig:contours}
\end{figure}

In Fig.~\ref{fig:contours}, situations in our model are summarized.
Six cases with different numbers of doubly charged scalars are considered with {$N_j=0,1,10,100,200$ and $300$}.
Here, we should mention an important issue.
As indicated in Fig.~\ref{fig:contours}, when $N_j$ is zero, more than {$10\sim20\,\text{TeV}$} is required in the effective trilinear coupling $\mu_{S k}$.
Such a large trilinear coupling {would immediately lead} to the violation of tree level unitarity in the scattering amplitudes including $\mu_{S k}$, e.g., $k^{++}k^{--} \to k^{++}k^{--}$ or $SS \to k^{++}k^{--}$ at around the energy $1\,\text{TeV}$, where the physics with our interest is spread.
Also, the vacuum is possibly threatened by the destabilization via the large trilinear coupling, which calls charge breaking minima.
To evade the problems, naively speaking, the value of $\mu_{S k}$ is less than {$1 \sim 5\,\text{TeV}$}.\footnote{
{In the case of MSSM with a light $\tilde{t}_1$ ($100\,\text{GeV}$), $A = A_t = A_b$, $\tan{\beta} \gg 1$, $m_A \gg M_Z$, $|\mu| \ll M_{\tilde{Q}}$ and $M_{\tilde{b}}$, the bound on the trilinear coupling $|A| \lesssim 5\,\text{TeV}$ was reported in Ref.~\cite{Schuessler:2007av}.}
}

\begin{figure}[t]
\begin{center}
\includegraphics[scale=0.5]{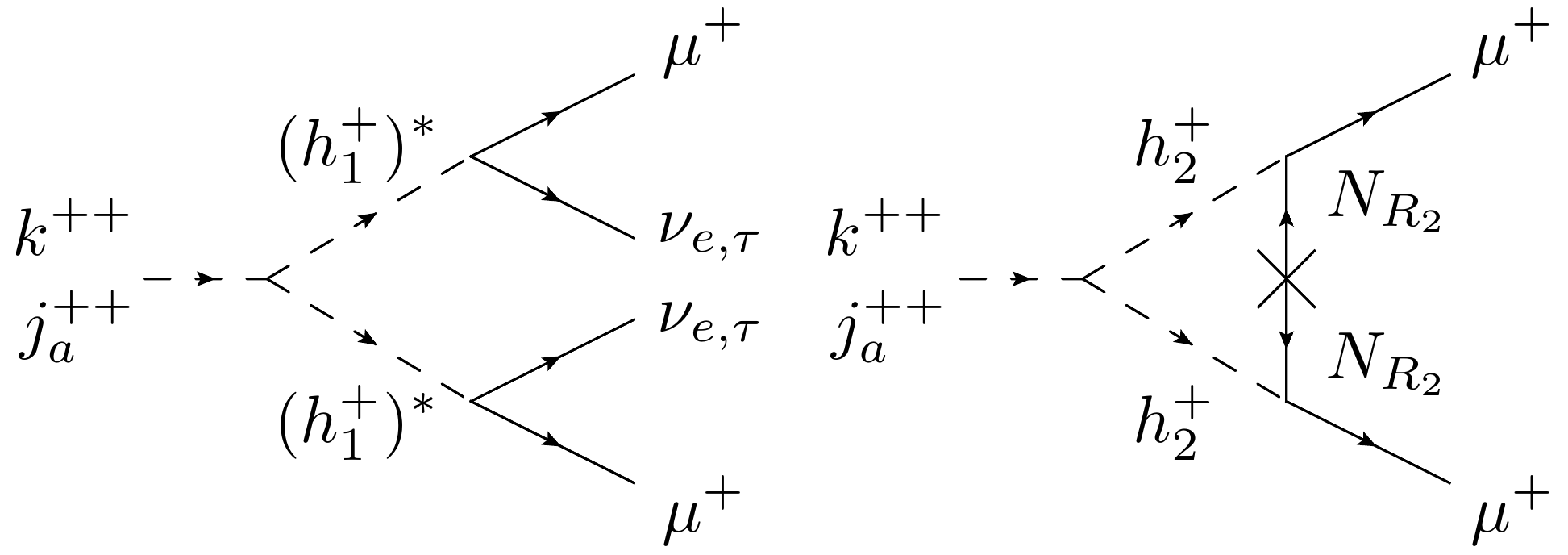}
   \caption{A schematic description for the decay patterns of $k^{++}$ or $j_a^{++}$ with two anti-muons in the final state.
{Here, $h_1^{+}$'s in the left diagram are off-shell particles.}}
   \label{fig:decay_diagram}
\end{center}
\end{figure}

Also, we consider the doubly charged singlet scalars produced via $p p \to \gamma^\ast/Z +X \to k^{++} k^{--} +X$.
Lower bounds at $95\%$ {C.L.} on $m_{k^{\pm\pm}}$ via the $8\,\text{TeV}$ LHC data were provided by the ATLAS group in Ref.~\cite{ATLAS:2014kca} as $374\,\text{GeV}$, $402\,\text{GeV}$, $438\,\text{GeV}$ when assuming a $100\%$ branching ratio to $e^{\pm} e^{\pm}$, $e^{\pm} \mu^{\pm}$, $\mu^{\pm} \mu^{\pm}$ pairs, respectively.
In our model, the doubly charged scalars can decay through the processes as shown in Fig.~\ref{fig:decay_diagram}, where $h_1^+$'s  are off shell since it should be heavy at least $3\,\text{TeV}.$
In the case of $k^{++}$ in $N_j = 0$, when the values of $\mu_{11}$ and $\mu_{22}$ are the same or similar, {from Eq.~(\ref{HP})}, the relative branching ratios between $k^{++} \to \mu^+ \mu^+ \nu_{i} \nu_{j}$ and $k^{++} \to \mu^+ \mu^+$ are roughly proportional to $(y_L)_{2i} (y_L)_{2j}$ and $((y_R)_{22})^2$.
As concluded in our previous work~\cite{Nishiwaki:2015iqa},
the absolute value of $(y_R)_{22}$ should be large as around $8 \sim 9$ to generate the observed neutrino properties, while a typical magnitude of {$(y_L)_{2i}$} is $0.5 \sim 1$.
Then, the decay branch $k^{++} \to \mu^+ \mu^+$ is probably dominant as $\sim 100\%$ and we need to consider the 8\,TeV bound seriously.
A simplest attitude would be to avoid to examine the shaded regions in Fig.~\ref{fig:contours}, which indicate the excluded parts in $95\%$ {C.L.} via the ATLAS 8\,TeV data with the assumption of $\B(k^{\pm\pm} \to \mu^\pm \mu^\pm) = 100\%$~\cite{ATLAS:2014kca}.

When one more doubly charged scalar $j_1^{++}$ ($N_j = 1$) {exists}, a detailed analysis is needed for precise bounds on $k^{\pm\pm}$ and $j_1^{\pm\pm}$.
Benchmark values are given in Fig.~\ref{fig:contours} by the vertical black {dotted} lines, which represent corresponding bounds on the universal physical mass $m_{k^{\pm\pm}}$ when we assume $\B(j^{\pm\pm}_a \to \mu^\pm \mu^\pm) = 100\%$ for all $j^{\pm\pm}_a$.
We obtain the $95\%$ {C.L.} lower {bounds} on the universal mass value $m_{k^{\pm\pm}}$ as
${\sim 500}\,\text{GeV}\,(N_j=1)$,
${\sim 660}\,\text{GeV}\,(N_j=10)$,
${\sim 900}\,\text{GeV}\,(N_j=100)$,
{${\sim 980}\,\text{GeV}\,(N_j=200)$, and
${\sim 1030}\,\text{GeV}\,(N_j=300)$},
respectively through the numerical simulations by {\tt MadGraph5\_aMC@NLO}~\cite{Alwall:2011uj,Alwall:2014hca} with the help of {\tt FeynRules}~\cite{Christensen:2008py,Alloul:2013bka,Degrande:2011ua} for model implementation.

The method which we adopt for evaluating the corresponding $95\%$ C.L. bounds with the assumption of $\B(j_a^{\pm\pm} \to \mu^\pm \mu^\pm) = 100\%$ for all $j^{\pm\pm}_a$, where more than one doubly charged scalars exist, is as follows.
When $N$ number of doubly charged scalars are {present}, the expected number of the total signal receives the multiplicative factor $N$.
Following this statement, we can estimate the bound on the universal mass $m_{k^{\pm\pm}}$ via the pair 
production cross section of a doubly charged scalar $k^{\pm\pm}$ (in $N=1$ case) though the sequence $p p \to \gamma^\ast/Z +X \to k^{++} k^{--} +X$.
The bound should correspond to the mass where the production cross section is $N$-times smaller than the benchmark value in $m_{k^{\pm\pm}} = 438\,\text{GeV}$, which is the $95\%$ C.L lower bound on $m_{k^{\pm\pm}}$ form the ATLAS 8\,TeV data~\cite{ATLAS:2014kca}.
We obtained the leading order cross section as $0.327\,\text{fb}$, which is fairly close to the ATLAS value, $0.357\,\text{fb}$ read from Fig.~4\,(c) of Ref.~\cite{ATLAS:2014kca}.
In calculation, we used {\tt CTEQ6L} proton PDF~\cite{Pumplin:2002vw} and set the renormalization and factorization scales as $2 m_{k^{\pm\pm}}$. 

Here, we point out an interesting possibility.
{From Eq.~(\ref{eq:additional_terms_effective}), if} $\lambda_{11}^{(1)} \langle {\Sigma_0^\ast} \rangle$ is quite larger than $\mu_{22}^{(1)}$, the pattern $j_1^{++} \to \mu^+ \mu^+ \nu_{i} \nu_{j}$ possibly becomes considerable, where we cannot reconstruct the invariant mass of the doubly charged scalar since missing energy exists in this decay sequence.
Then, significance {for exclusion} would be dropped and we could relax the bound on $m_{j_1^{\pm\pm}}$ to some extent.
An extreme case is with a nonzero $\lambda_{11}^{(1)} \langle \Sigma_0^\ast \rangle$ and $\mu_{22}^{(1)} = 0$, where the branching ratio of $j_1^{++} \to \mu^+ \mu^+$ becomes zero at the one loop level and the significance takes the lowest value, which is the best for evading the 8\,TeV LHC bound.
Also in this situation, no additional contribution to the neutrino mass matrix exists and the original successful structure is not destroyed. 
Similar discussions are applicable when $N_j$ is more than one.

When we assume $100\%$ branching fractions in $j_a^{++} \to \mu^{+} \mu^{+}$ for all $j_a^{++}$,
the common trilinear coupling $\mu_{S k}$ should be larger than
{$\sim 10\,\text{TeV}\,(N_j=0)$,
$\sim 8\,\text{TeV}\,(N_j=1)$,
$\sim 3\,\text{TeV}\,(N_j=10)$,
less than $1\,\text{TeV}\,(N_j=100,\,200,\,300)$,
to obtain a reasonable amount of the production cross section with taking into account of the $\pm 20\%$ theoretical error in cross section as suggested by Fig.~\ref{fig:contours}}.
As mentioned, large trilinear couplings $\lambda_{11}^{(a)} \langle \Sigma_0^\ast \rangle$ can help us to alleviate the 8\,TeV bound.

\begin{figure}[t]
\begin{center}
\includegraphics[scale=0.6]{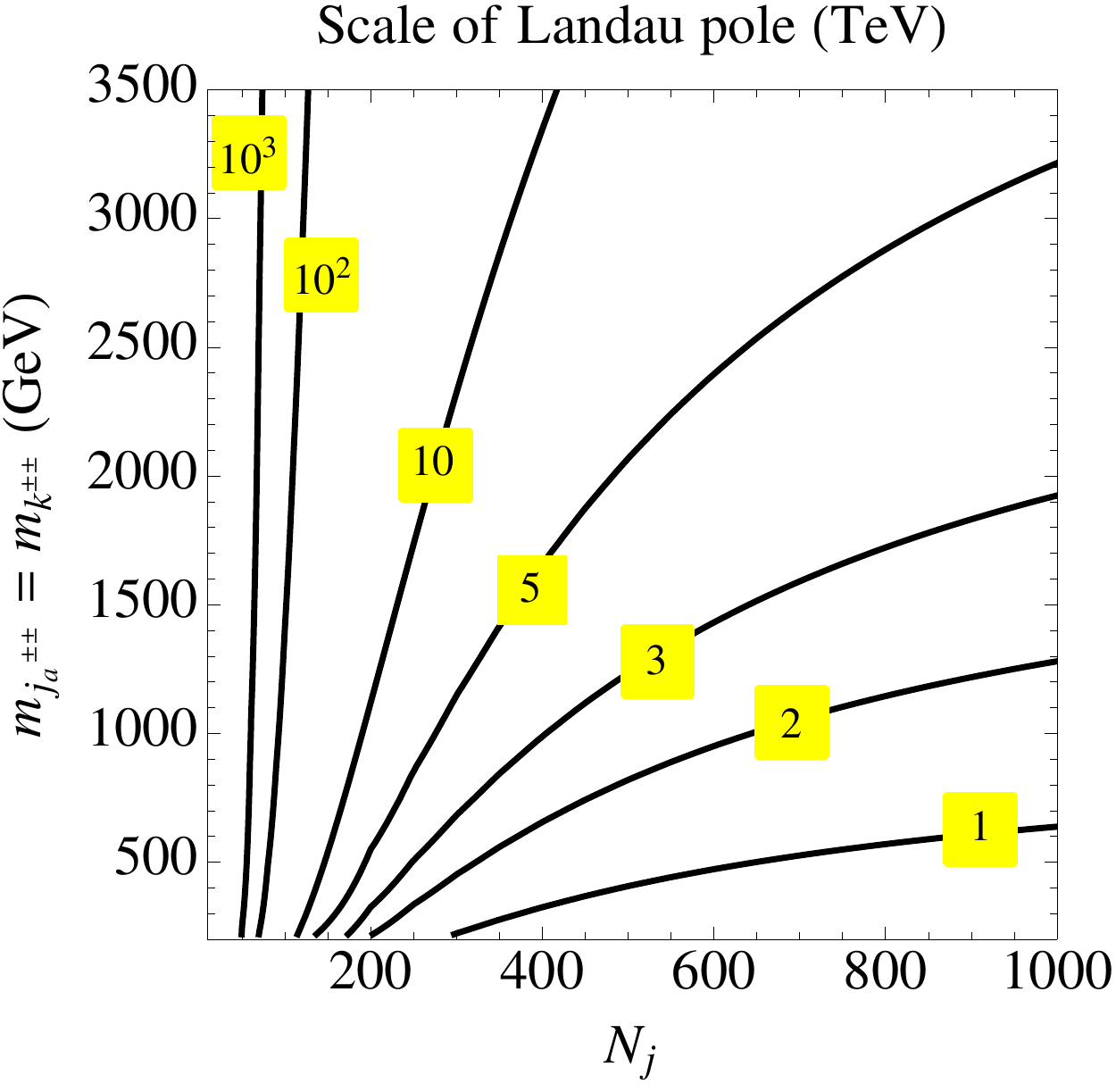}
   \caption{Positions of ``Landau pole'' defined as $g_{Y} (\mu) = 4\pi$}
   \label{fig:Landaupole}
\end{center}
\end{figure}

%
Another theoretical bound is reasonably expected when, as in the present situation, many new particles with nonzero gauge charges are introduced around $1\,\text{TeV}$.
The presence of multiple doubly charged $SU(2)_{L}$ singlet scalars deforms the energy evolution of the $U(1)_{Y}$ gauge coupling $g_{Y}$ as
\al{
\frac{1}{g_{Y}^{2}(\mu)} = \frac{1}{g_{Y}^{2}(m_{\text{input}})} -
\frac{b_{Y}^{\text{SM}}}{16\pi^{2}} \log\left( \frac{\mu^{2}}{m_{\text{input}}^{2}} \right) -
\theta(\mu - m_{\text{threshold}})
\frac{\Delta b_{Y}}{16\pi^{2}} \log\left( \frac{\mu^{2}}{m_{\text{threshold}}^{2}} \right),
}
where $b_{Y}^{\text{SM}} = 41/6$, $\Delta b_{Y} = 4/3 (N_{j} + 1)$, and we implicitly assume the relation $m_{\text{input}} \, (=m_{Z}) < m_{\text{threshold}} \, (= m_{k^{\pm\pm}} = m_{j_{a}^{\pm\pm}})$.
As a reasonable criterion, we require that the theory is still not drastically strongly-coupled {within the LHC reach $\sim 10$ TeV}.\footnote{
{We note that} measurements of running electroweak couplings put bounds on additional contributions to the beta functions of the $SU(2)_{L}$ and $U(1)_{Y}$ gauge couplings~\cite{Alves:2014cda} {even though} the work~\cite{Alves:2014cda} did not survey the parameter range which is relevant for our discussion.
Similar discussions have been done in the QCD coupling~\cite{Becciolini:2014lya,Bae:2016xni}{, which is basically irrelevant in our case.}
}
Positions of ``Landau pole'' $\mu$, which is defined as $g_{Y} (\mu) = 4\pi$, are calculated with ease as functions of $N_{j}$ and $m_{\text{threshold}} \, (= m_{k^{\pm\pm}} = m_{j_{a}^{\pm\pm}})$ as shown in Fig.~\ref{fig:Landaupole}.
Now, we recognize that under the criterion, the case with $N_{j} \lesssim 100$ {is not restricted in the sense that}
the bound via the ``Landau pole'' is much weaker than the phenomenological requirement $m_{k^{\pm\pm}} \, (= m_{j_{a}^{\pm\pm}}) \gtrsim 375\,\text{GeV}$ (for preventing the decays $S \to k^{++} k^{--},\, j_{a}^{++} j_{a}^{--}$).
On the other hand when $N_{j}$ is rather larger than $100$, meaningful bounds are expected from Fig.~\ref{fig:Landaupole}.
For example, when $N_{j} = 200$ (300), $m_{k^{\pm\pm}} \, (= m_{j_{a}^{\pm\pm}})$ should be greater than $\sim 1.1\,\text{TeV}$ ($\sim 2.2\,\text{TeV}$).

There also arises a largish loop contribution to the universal trilinear coupling $\mu_{Sk} \, (= \mu_{S j_{a}})$ as
\al{
\mu_{Sk} \to \mu_{Sk} \, (1 + c \, \delta),\quad
\delta = \frac{N_j+1}{16\pi^2} \times \left(\frac{\mu_{S k}}{m_{k^{\pm\pm}}}\right)^2.
\label{eq:trilinear_correction}
}
A convenient parameter, $c \lsim 1$, encapsulates the effects from all higher order contributions. Precise determination of $c$ is beyond the scope of this paper thus, instead, we show the cases with $c=0.01, 0.1$ and $1$ as benchmarks (see Fig \ref{fig:contours}).
It is easily noticed that the loop-induced value could dominate over the tree level value unless $c \, \delta < 1$, or equivalently $\mu_{S k}/m_{k^{\pm\pm}} < 4\pi/\sqrt{c \, (N_j+1)}$.
This may affect the convergence of the multi-loop expansion even though the theory is still renormalizable.\footnote{{One should note, however, that $c\,\delta<1$ is not absolute requirement for a consistent theory. See e.g. \cite{Kanemura:2004mg} where a loop-induced value overwhelms the tree-level counterpart in the context of two Higgs doublet model.}
}

Unfortunately when $N_j$ is only a few, explaining the diphoton excess is not consistent since the value of $\mu_{S k}$ is too large and tree level unitarity is violated.
This problem is avoided when $N_j \gtrsim 10$, whereas the evolution of $g_{Y}$ through renormalization group effect puts additional bounds on $m_{k^{\pm\pm}} \, (= m_{j_{a}^{\pm\pm}})$ when $N_j \gtrsim 100$.  {The preferred parameter would be further constrained by $c\,\delta<1$ as in Fig. \ref{fig:contours}.} 
In conclusion, we can explain the 750\,\text{GeV} excess consistently even when $\B(j^{\pm\pm}_a \to \mu^\pm \mu^\pm) = 100\%$ for all $j_a^{\pm\pm}$.

\subsubsection{Case 2: with mass mixing
\label{sec:withmixing}}

%
In this section, we investigate the situation when the mass mixing between  $S$ and $\Phi$ are allowed.
At first, we {phenomenologically introduce} the mixing angle $\beta$ as,
\al{
\begin{pmatrix} \phi \\ S \end{pmatrix}
=
\begin{pmatrix}
c_{\beta} & s_{\beta} \\ -s_{\beta} & c_{\beta}
\end{pmatrix}
\begin{pmatrix} h \\ S' \end{pmatrix},
	\label{eq:scalar_mixing_formulation}
}
where we use the short-hand notations, $c_{\beta} \equiv \cos{\beta}$, $s_{\beta} \equiv \sin{\beta}$, and express the observed $125\,\text{GeV}$ and $750\,\text{GeV}$ scalars (mass eigenstates) by $h$ and $S'$, respectively.
We assume the following effective interactions among scalars
\begin{eqnarray}
\Delta \mathcal{V}_{\text{eff}} =
\frac{1}{2} m_{h}^{2} h^{2} + \frac{1}{2} m_{S'}^{2} S'^{2} +
\mu_{Sk} S |k^{++}|^{2} + \mu_{Sj_{a}} S |j_{a}^{++}|^{2} + \hat{\mu}_{S \Phi} S|\Phi|^{2} +
\hat{\lambda}_{S\Phi} S^{2} |\Phi|^{2},
\end{eqnarray}
where $m_{h}$ and $m_{S'}$ represent the mass eigenvalues $125\,\text{GeV}$ and $750\,\text{GeV}$;
$\mu_{Sk}$ and $\mu_{Sj_{a}}$ are effective trilinear couplings as defined in Eq.~(\ref{eq:effective_scalar_trilinear}), where the contents of them are not important in this study.
{
We note that we safely ignore the terms {$\phi |k^{++}|^{2}$ and $\phi |j_{a}^{++}|^{2}$} since these terms originate from
the gauge-invariant interactions {$|\Phi|^{2} |k^{++}|^{2}$ and $|\Phi|^{2} |j_{a}^{++}|^{2}$}, where effective trilinear couplings of them are small compared with $\mu_{Sk}$ and $\mu_{Sj_{a}}$.}
Because of the mixing in Eq.~(\ref{eq:scalar_mixing_formulation}), the terms $h |k^{++}|$ and $h |j_{a}^{++}|$ are induced and can affect the signal strength of $h$.

The $S'$-$h$-$h$ interaction may be also introduced via the interaction Lagrangian:
\al{
\frac{1}{2} \mu_{S'h} S' h^{2}\quad
\text{with}\quad
\mu_{S'h} \equiv m_{S' h} \left[ c_{\beta}^{3} - 2 c_{\beta} s_{\beta}^{2} \right],
	\label{eq:form_of_trilinear}
}
where $m_{S' h}$ represents a mass scale and the mixing factor could be determined via the gauge invariant term $S |\Phi|^{2}$.\footnote{
When $\langle S \rangle = 0$, the scale of $m_{S' h}$ is determined through the two mass eigenvalues and the mixing angle $\beta$ as
\al{
m_{S' h} = \left( \frac{m_{S'}^{2} - m_{h}^{2}}{v} \right) \sin(2\beta),
	\label{eq:m_correspondence}
}
since the mass mixing term $S\phi$ and the three point vertex $S \phi^{2}$ have the unique common origin $S |\Phi|^{2}$.
{Plots in this situation are provided in Appendix~\ref{sec:app:plots}.}
}

\begin{figure}[t]
\begin{center}
\includegraphics[width=0.49\columnwidth]{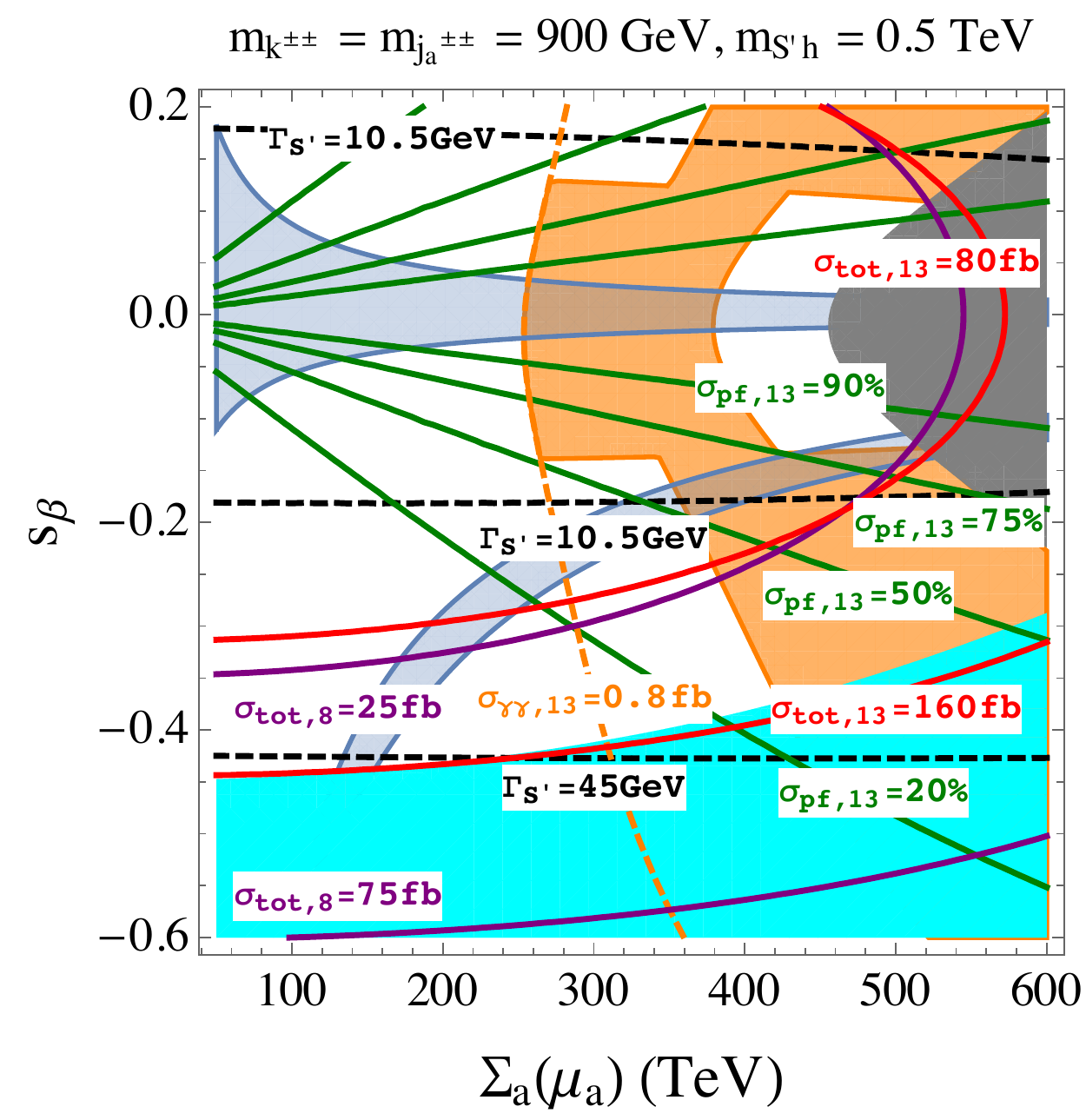}\
\includegraphics[width=0.49\columnwidth]{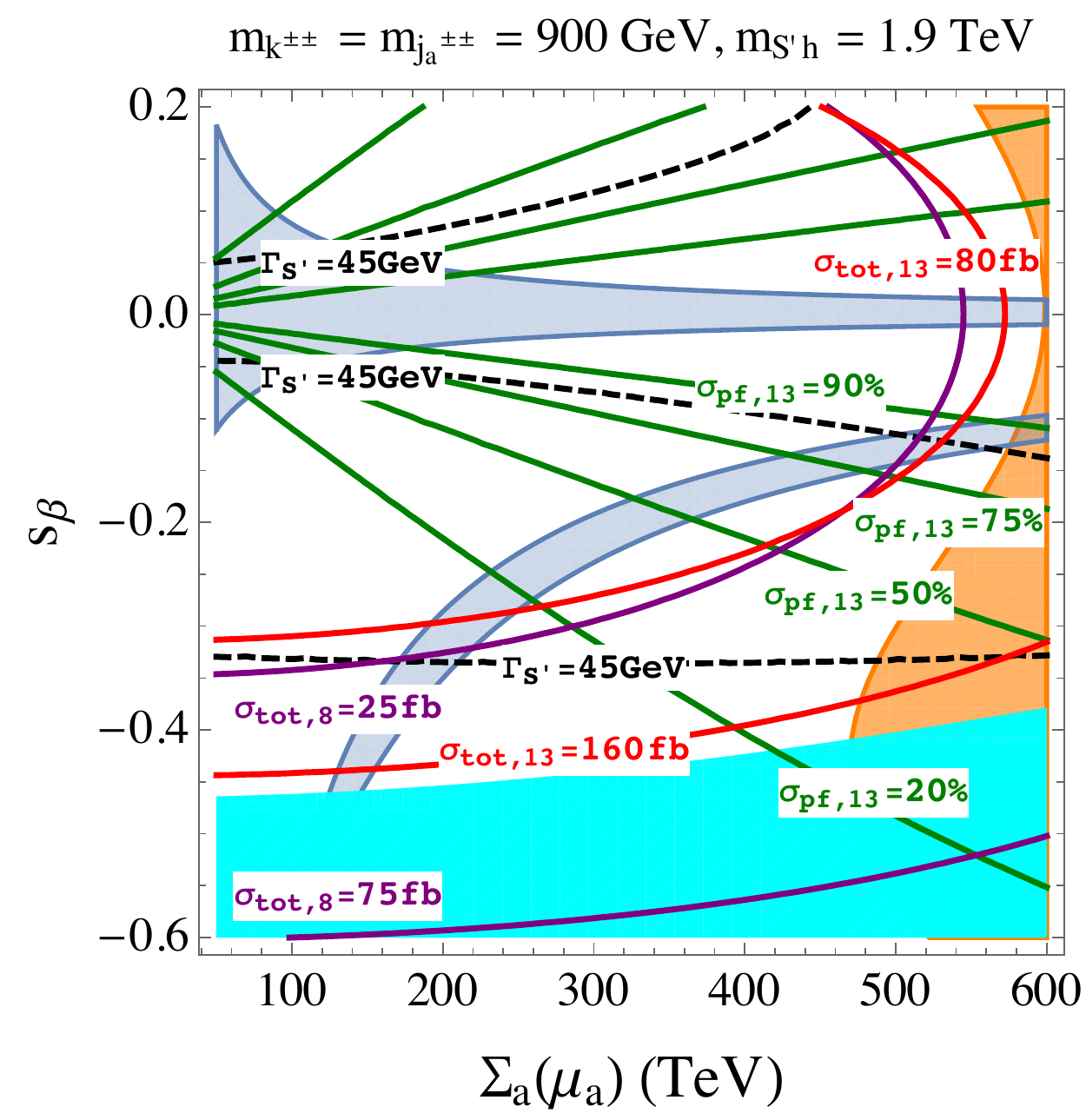}
\caption{Allowed ranges of the parameters $\{ \sum_{a}(\mu_{a}), s_{\beta} \}$ are shown {in the choice of the mass of the degenerated doubly charged scalars ($m_{k^{\pm\pm}} \, [= m_{j_{a}^{\pm\pm}}] = 900\,\text{GeV}$)} and {two different choices of $m_{S'h}$ ($0.5\,\text{TeV}$ [left panel] and $1.9\,\text{TeV}$ [right panel])}.
The light blue regions represent $2\sigma$ allowed regions of $125\,\text{GeV}$ Higgs signal strengths, while the orange regions suggest the areas where the $750\,\text{GeV}$ excess is suitably explained.
The gray/cyan regions are excluded in $95\%$ C.L.s by the ATLAS $8\,\text{TeV}$ results for $S' \to \gamma\gamma/ZZ$.
For better understanding, several contours for the total width of $S'$ ($\Gamma_{S'}$),
total production cross sections at $\sqrt{s} = 8/13\,\text{TeV}$ ($\sigma_{\text{tot},8/13}$),
and the percentage of the production through the photon fusion at $\sqrt{s} = 13\,\text{TeV}$ ($\sigma_{\text{pf},13}$) are illustrated. 
}
   \label{fig:allowedregion}
\end{center}
\end{figure}

\begin{figure}[t]
\begin{center}
\includegraphics[width=0.45\columnwidth]{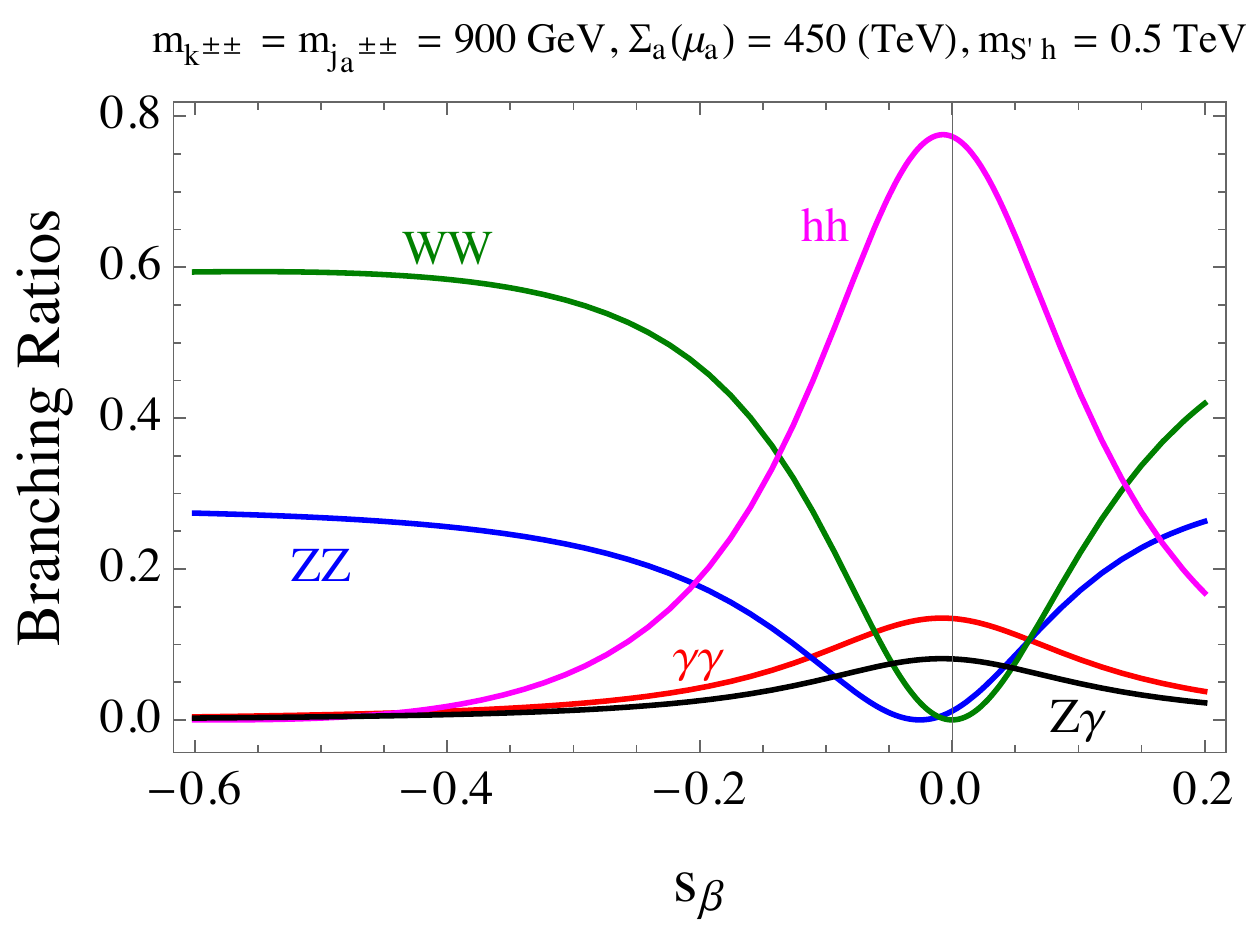}\
\includegraphics[width=0.45\columnwidth]{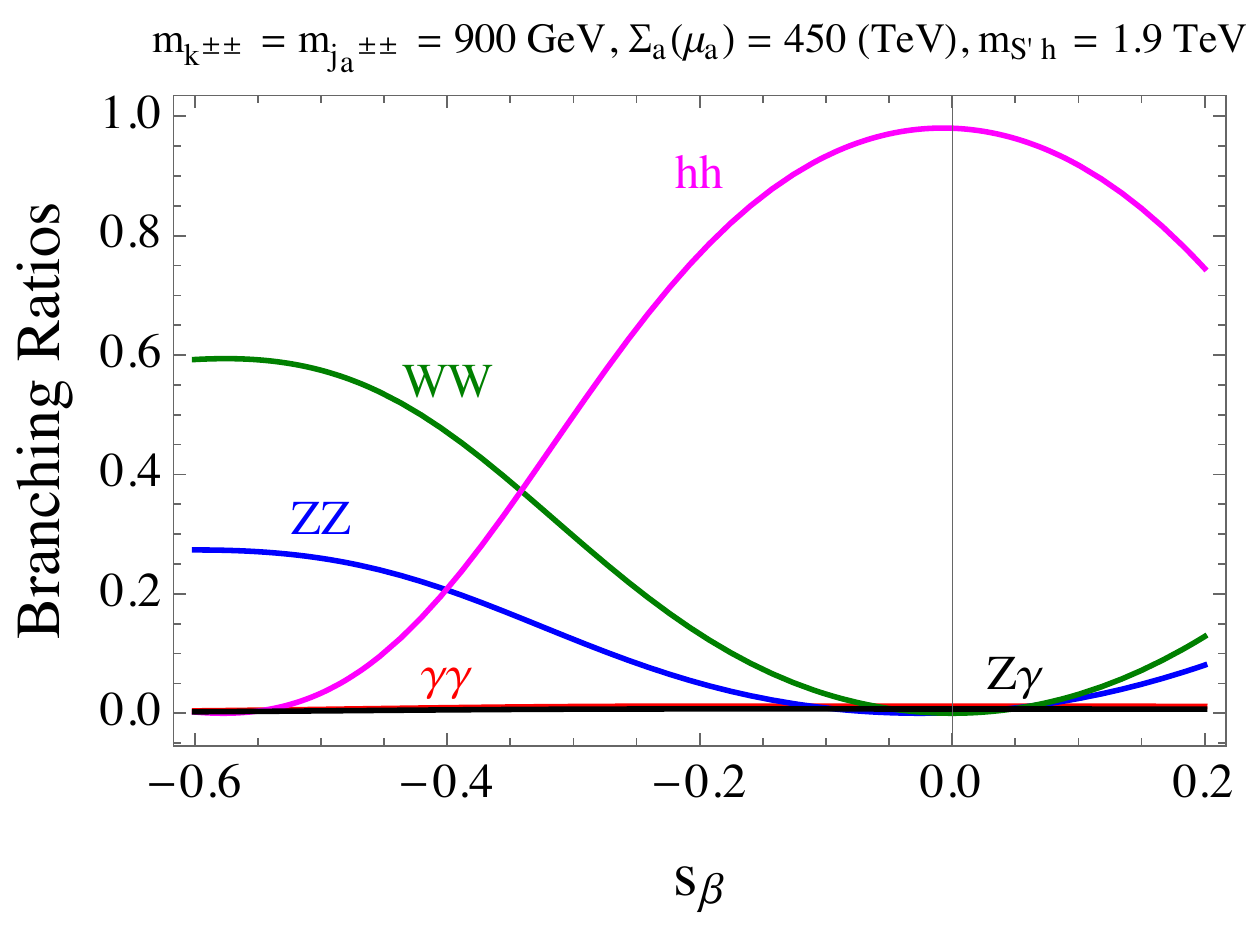}
\caption{Relevant branching ratios of $S'$ in the two configurations in Fig.~{\ref{fig:allowedregion}} are shown.
Here, values of $\Sigma_{a} (\mu_{a})$ are suitably fixed as typical digits in the corresponding allowed regions.
}
   \label{fig:Sprime_Brs_1}
\end{center}
\end{figure}

A signifiant distinction from the previous no-mixing case is that the $750\,\text{GeV}$ scalar can couple to the SM particles through the mixing effect.
The inclusive production cross section at the LHC is deformed as
\al{
\sigma(pp \to S' + X) \simeq (\sigma^{\text{ggF}}_{pp \to H_{750\,\text{GeV}}^{\text{SM}}} + \sigma^{\text{VBF}}_{pp \to H_{750\,\text{GeV}}^{\text{SM}}}) s_{\beta}^{2} + \sigma_{pp \to S'}^{\text{pf}},
}
where $\sigma^{\text{ggF}}_{pp \to H_{750\,\text{GeV}}^{\text{SM}}}$ and $\sigma^{\text{VBF}}_{pp \to H_{750\,\text{GeV}}^{\text{SM}}}$ represent the inclusive production cross section of the SM-like Higgs boson with $750\,\text{GeV}$ mass through the gluon fusion and vector boson fusion processes, respectively.
$\sigma_{pp \to S'}^{\text{pf}}$ shows a corresponding value through the photon fusion in Eq.~(\ref{eq:photoproduction_total}).
We adopt the following digits in~\cite{Altmannshofer:2015xfo,Dittmaier:2011ti,Dittmaier:2012vm,Heinemeyer:2013tqa},
\al{
\sigma^{\text{ggF}}_{pp \to H_{750\,\text{GeV}}^{\text{SM}}} =
\begin{cases}
156.8\,\text{fb} & \text{at } \sqrt{s} = 8\,\text{TeV} \\
590\,\text{fb} & \text{at } \sqrt{s} = 13\,\text{TeV}
\end{cases},\quad
\sigma^{\text{VBF}}_{pp \to H_{750\,\text{GeV}}^{\text{SM}}} =
\begin{cases}
50\,\text{fb} & \text{at } \sqrt{s} = 8\,\text{TeV} \\
220\,\text{fb} & \text{at } \sqrt{s} = 13\,\text{TeV}
\end{cases}, \\[4pt]
\Gamma_{\text{tot}}(H_{750\,\text{GeV}}^{\text{SM}}) = 247\,\text{GeV},\quad
\B(H_{750\,\text{GeV}}^{\text{SM}} \to WW) = 58.6\%,\quad
\B(H_{750\,\text{GeV}}^{\text{SM}} \to ZZ) = 29.0\%.
}
{Part of} relevant partial decay widths are written down as
\al{
\Gamma_{S' \to WW} &= \Gamma_{H_{750\,\text{GeV}}^{\text{SM}} \to WW} \, s_{\beta}^{2}, \\
\Gamma_{S' \to hh} &\sim \frac{(\mu_{S' h})^{2}}{32\pi m_{S'}} \sqrt{1 - \frac{4 m_{h}^{2}}{m_{S'}^{2}}},
}
The total width takes the form
\al{
\Gamma_{\text{tot}}(S') \sim {\left[ \Gamma_{\text{tot}}(H_{750\,\text{GeV}}^{\text{SM}}) -
\Gamma_{H_{750\,\text{GeV}}^{\text{SM}} \to ZZ} \right] s_{\beta}^{2}
+ \Gamma_{S' \to \gamma\gamma}
+ \Gamma_{S' \to Z\gamma}
+ \Gamma_{S' \to ZZ}}
+ \Gamma_{S' \to hh},
}
where the minuscule parts $\B({H_{750\,\text{GeV}}^{\text{SM}} \to \gamma\gamma}) = 1.79 \times 10^{-5}\%$, $\B({H_{750\,\text{GeV}}^{\text{SM}} \to Z\gamma}) = 1.69 \times 10^{-4}\%$, {and $\B(H_{750\,\text{GeV}}^{\text{SM}} \to gg) = 2.55 \times 10^{-2}\%$}~\cite{Heinemeyer:2013tqa} {could be safely neglected.}
{Here, $\Gamma_{S' \to \gamma\gamma}$, $\Gamma_{S' \to Z\gamma}$ and $\Gamma_{S' \to ZZ}$ describe decay widths at one loop level, where the multiple doubly charged scalars propagating in the loops.}
When we take the limit $s_{\beta} \to 0$, they are reduced to Eqs.~(\ref{eq:StoVV_1})-(\ref{eq:StoVV_3}).
{Explicit forms of these widths are summarized in {Appendix~\ref{sec:app:width}}.}

In Fig.~{\ref{fig:allowedregion}}, prospects are widely discussed {in the choice of the mass of the degenerated doubly charged scalars ($m_{k^{\pm\pm}} \, [= m_{j_{a}^{\pm\pm}}] = 900\,\text{GeV}$)} and {two different choices of $m_{S'h}$ ($0.5\,\text{TeV}$ [left panel] and $1.9\,\text{TeV}$ [right panel])}.
First, we emphasize that the $125\,\text{GeV}$ Higgs $h$ couples to the doubly charged scalars through the mixing in Eq.~(\ref{eq:scalar_mixing_formulation}) in the present setup.
As in Ref.~\cite{Nishiwaki:2015iqa}, we take the results at $\sqrt{s} = 7 \text{ and } 8\,\text{TeV}$ of the five Higgs decay channels reported by the ATLAS and the CMS experiments into consideration, which are $h \to \gamma\gamma$, $h \to ZZ$, $h \to WW$, $h \to b \bar{b}$, $h \to \tau^{+} \tau^{-}$~\cite{Aad:2014eva,Aad:2014eha,Aad:2014xzb,ATLAS-CONF-2014-061,ATLAS:2014aga,Khachatryan:2014jba}, and calculate a $\chi^{2}$ variables for estimating $2\sigma$ allowed ranges of the parameter space, which are depicted in light blue color.\footnote{
{The original model contains invisible channels in the $125\,\text{GeV}$ Higgs boson due to the existence of a dark matter candidate and a Nambu-Goldstone boson from the spontaneous breaking of a global $U(1)$.
We ignore the invisible widths in the global fit for simplicity.}
}
Here, we find two types of allowed regions with and without including $s_{\beta} = 0$, which correspond to the cases with and without accidental cancellation between SM contributions and the new contributions through the mixing, respectively.

The orange regions suggest the $2\sigma$-favored areas with taking account of the $20\%$ theoretical error in the present way for photon-fusion production cross section summarized in Eq.~(\ref{eq:bestfit_range_2}).
Here, we use the values in the cases of $\Gamma/m \to 0$ and $\Gamma/m = 6\%$ for the regions $\Gamma/m < 1\%$ and $\Gamma/m \geq 1\%$ for an illustration, respectively.
The gray/cyan regions are excluded in $95\%$ C.L.s by the ATLAS $8\,\text{TeV}$ results for $S' \to \gamma\gamma/ZZ$.
For better understanding, several contours for the total width of $S'$ ($\Gamma_{S'}$),
total production cross sections at $\sqrt{s} = 8/13\,\text{TeV}$ ($\sigma_{\text{tot},8/13}$),
and the percentage of the production through the photon fusion at $\sqrt{s} = 13\,\text{TeV}$ ($\sigma_{\text{pf},13}$) are illustrated.
Relevant branching ratios of $S'$ are shown in Fig.~{\ref{fig:Sprime_Brs_1}} for the {two} configurations in Fig.~{\ref{fig:allowedregion}}.

Now, we focus on two types of consistent solutions around $s_{\beta} \simeq 0$ and {$s_{\beta} \simeq -0.15$}. 
Physics in the situation $s_{\beta} \simeq 0$ is basically the same with the previous ``Case 1'' without mass mixing effect, where the total decay width is small, concretely less than $1\,\text{GeV}$.
On the other hand when {$s_{\beta} \simeq -0.15$, partial widths of decay branches which are opened by a nonzero value of $s_{\beta}$ become sizable and expected values of the total width can become, interestingly, near $10.5\,\text{GeV}$ or $45\,\text{GeV}$, which are the latest $13\,\text{TeV}$ best fit value of the CMS and ATLAS group, respectively.}

Finally, we briefly comment on tree level unitarity.
When we consider $m_{k^{\pm\pm}} \, [= m_{j_{a}^{\pm\pm}}] = 900\,\text{GeV}$, the bound via tree level unitarity is relaxed in both of $s_{\beta} \simeq 0$ and {$s_{\beta} \simeq -0.15$}.
{However, with a large value of the universal trilinear coupling in 3 to 6 TeV range,  $c \, \delta < 1$ is achieved only if $c\ll1$ when $\B(j^{\pm\pm}_a \to \mu^\pm \mu^\pm) = 100\%$ for all $j^{\pm\pm}_a$, which may requests further model building efforts.}

\section{Conclusion and discussion
\label{sec:summary}}

In this paper, we investigated a possibility for explaining the recently announced 750\,GeV diphoton excess by the ATLAS and the CMS experiments at the CERN LHC in the context of loop induced singlet production and decay through photon fusion.
When a singlet scalar $S$, which is a candidate of {the resonance particle}, couples to {doubly} charged particles, we can obtain a suitable amount of the cross section of {$p p \to S+X \to \gamma\gamma+X$} without {introducing} a tree-level production of $S$.
In three-loop radiative neutrino models, $SU(2)_L$ singlet multiple doubly charged scalars are  introduced such that the $S$-$\gamma$-$\gamma$ vertex is radiatively generated and enhanced.
When we consider such type of doubly charged scalar(s), the branching ratio $\B(S \to \gamma\gamma)$ is {uniquely} fixed as $\simeq 60\%$ by quantum numbers {when $S$ is a mass eigenstate.}
Constraints from $8\,\text{TeV}$ LHC data are all satisfied.

A fascinating feature in the single $S$ production through photon fusion is that the value of $\B(S \to \gamma\gamma)$ as well as $\Gamma_S$ determines the production cross section, as shown in Eqs.~(\ref{eq:photoproduction_total}) and (\ref{eq:photoproduction_togammagamma}).
With the branching fraction to diphoton $S \to \gamma\gamma \simeq 60\%$ (see Sec.~\ref{sec:withoutmixing}), {when we take `wide width' scenario with $\Gamma/m \sim 6\%$,}
the expected cross section to diphton is too large.
However, in `narrow width scenario' with $\Gamma_{S} = 62.9\,\text{MeV}$, it is nicely fit to the best fit value for the inclusive cross section of $2\,\text{fb}$. We also note that the width is close to the $8 + 13\,\text{TeV}$ best-fit value announced by the CMS group ($105\,\text{MeV}$) [see App.~\ref{sec:app:plots}].
This is an informative prediction of our present scenario which should be tested in the near future.
Also the relative strengths of the one loop induced partial decay widths are insensitive to $N_j$ as shown in Eq.~(\ref{eq:oneloopGamma_ratios}) when the mixing effect between $S$ and the Higgs doublet $\Phi$ is negligible.
This universality is a remarkable property of our scenario and this relation can be tested when more data would be available.

When $S$ and the Higgs doublet $\Phi$ can mix, some distinctive and interesting features are found. In the first thought, only a small mixing $\sin \beta \ll 1$ is allowed to circumvent drastic modifications to $125\,\text{GeV}$ Higgs signal strengths but we could see another interesting region of parameter space with $\sin \beta \simeq -0.15$, where the $750\,\text{GeV}$ excess can be explained consistently within `wide width' scenario (see Sec. \ref{sec:withmixing}).
{However a big part of the parameter space, especially in the case with the scalar mixing, would lie outside of $c\,\delta <1$ region, which requires $c\ll 1$ for a viable model.}

Finally, we discuss further extensions of the model and other phenomenological issues.
\begin{itemize}
\item
A possible extension of the present direction is to introduce {$N_S$ number of $SU(2)_L$ singlet scalars, $(S=S_1,S_2,\cdots,S_{N_S})$,}  without hypercharge in the theory.
If the masses {of the scalars are almost degenerate to $750\,\text{GeV}$,} the current experiment {may not be able to} detect the multi bumps {so that} they would look as a single bump as we face.
{The total cross section, then, is enhanced by the multiplicative factor of $N_S^2$ as
\begin{eqnarray}
\sigma_{\rm tot}(pp\to \gamma\gamma+X) 
\approx N_S^2 \sigma(pp \to S+X\to \gamma\gamma+X).
\end{eqnarray}
}
%
\item
Another possible extension is that we also introduce the singly charged scalars {$\tilde{h}^{\pm}_{1,2}$} which hold the same quantum numbers as $h_{1,2}^\pm$ and has the same interaction with $j_a^{\pm\pm}$ as $h_{1,2}^\pm$ do with $k^{\pm\pm}$.
In such a possibility, contributions to the neutrino mass matrix are enhanced and we can reduce the value of the large coupling required for a consistent explanation in the original model, especially in $(y_R)_{22}$.
See Appendix for details.
\item
{The triple coupling of the Higgs boson could be enhanced in our case that may activate} the strong first order phase transition, which is a necessity for realizing the electroweak baryogenesis scenario~\cite{Kanemura:2004ch}.
{In such a case, radiative seesaw models can explain not only neutrino mass and dark matter but also baryon asymmetry of the Universe.}
\item
{The decays $k^{\pm\pm} \to \ell^{\pm} \ell^{\pm}$ and $j^{\pm\pm}_a \to \ell^{\pm} \ell^{\pm}$ provide very clean signatures.
The 13\,TeV LHC would {be expected to} replace the current bound on the universal mass, {e.g., $m_{k^{\pm\pm}}> 438\,\text{GeV}$ {when $\B(k^{\pm}/j_a^{\pm} \to \mu^{\pm} \mu^{\pm}) = 100\%$ for all the doubly charged scalars}, from the 8\,TeV data}~\cite{ATLAS:2014kca} soon.
An important feature recognized from Fig.~\ref{fig:contours} is that when $N_j$ is not so large as around {$10$}, only light doubly charged scalars are consistent with the bound from tree level unitarity.
Such possibilities would be exhaustively surveyed and eventually confirmed or excluded in the near future.
On the other hand, when $N_j$ is large as around {$10$}, from Fig.~\ref{fig:contours}, more than $\sim 700\,\text{GeV}$ doubly charged scalars can exist with holding tree level unitarity.
Such heavy particles require a suitable amount of integrated luminosity for being tested in colliders.
In other words, such possibilities are hard to be discarded in the near future.
}
\item
It might be worth mentioning the discrimination between our model {discussed here} and the other {well known} radiative models{, namely, Zee model~\cite{Zee:1980ai} at the one-loop level, Zee-Babu model~\cite{Zee:1985id,Babu:1988ki} at the two-loop level, Kraus-Nasri-Trodden (KNT) model~\cite{Krauss:2002px}, {Aoki-Kanemura-Seto (AKS)} model~\cite{Aoki:2008av,Aoki:2011zg}, and Gustafsson-No-Rivera (GNR) model~\cite{Gustafsson:2012vj} at the three-loop level.}
Essentially, any model that include{s} isospin singlet charged bosons potentially explain the 750\,GeV diphoton excess along the same {way} as discussed in this {paper}.
{Among those, three-loop models have natural DM candidates by construction, which we regard as a phenomenological big advantage. Our model shares this virtue.} 
On the other hand, in view of the charged-boson, {our model and also the GNR model include doubly charged particles.}
{From the currently available data, it is not possible to distinguish the effect of a singly charged scalar from a double charged scalar.}
However, we {still see that} a doubly charged boson is in favor of the explanation of the 750 GeV diphoton excess {simply because of the enhanced diphoton coupling}.
\item
As we discussed before, $k^{\pm\pm}$ decays to $\mu^\pm \mu^\pm$ with an almost $100\%$ branching fraction, {distinctively from other models, e.g., Zee-Babu model,} due to the {large} coupling ${(y_R)_{22}} {\gsim 2\pi}$, which is required to realize the observed neutrino {data} in our setup consistently.
\end{itemize}

\bigskip

\textbf{{Note Added:}}
In the recent update in ICHEP 2016 (on 5th August 2016) after we submitted this manuscript to PTEP, which includes the analyzed data accumulated in 2016 (ATLAS: $15.4\,\text{fb}^{-1}$, CMS: $12.9\,\text{fb}^{-1}$), the 750\,GeV diphoton signal now turns out to be statistically disfavored~\cite{ATLAS:2016eeo,CMS:2016crm}.
Nevertheless, we are still motivated to study the diboson resonance which may show up in a higher energy domain\footnote{It is suggested that additional jet activity could provide a useful handle to understand the
underlying physics of heavy resonance in Ref.~\cite{Fuks:2016qkf}.} and the generic results in this paper would be useful in the future in any case.

\section*{Acknowledgements}

SK, KN, YO, and SCP thank the workshop, Yangpyung School 2015, for providing us an opportunity to initiate this collaboration.
We are grateful for Eung~Jin~Chun, Satoshi~Iso, Takaaki~Nomura and Hiroshi~Yokoya for fruitful discussions.
KN thanks Koichi Hamaguchi for useful comments when the first revision had been prepared.
SK was supported in part by Grant-in-Aid for Scientific Research, The Ministry of Education,
Culture, Sports, Science and Technology (MEXT), No.~23104006, and Grant H2020-MSCA-RISE-
2014 No.~645722 (Non-Minimal Higgs).
This work is supported in part by NRF Research No. 2009-0083526 (YO) of the Republic of Korea.
SCP is supported by the National Research Foundation of Korea (NRF)
grant funded by the Korean government (MSIP) (No. 2016R1A2B2016112) and (No. 2013R1A1A2064120).
This work was supported by IBS under the project code, IBS-R018-D1 for RW.

\appendix
\section*{Appendix}


\section{Brief review on the original model
\label{sec:app:review}}

Here, we briefly summarize features in the model discussed in~\cite{Nishiwaki:2015iqa}.
\begin{enumerate}
\renewcommand{\labelenumi}{(\alph{enumi})}
\item In this model, the sub-eV neutrino masses are radiatively generated at the three loop level with the loop suppression factor $1/(4\pi)^6$.
{In such a situation}, a part of couplings, including scalar trilinear couplings, contributing to the neutrino matrix tends to be close to unity.
\item When a scalar trilinear coupling is large, they can put a negative effect on scalar quartic couplings at the one loop level, which threatens the stability of the vacuum.
\item The doubly charged scalar $k^{\pm\pm}$ is isolated from the charged lepton {at the leading order} under the assignment of the global $U(1)$ charges summarized in Tab.~\ref{tab:1}.
Then, the charged particle does not contribute to lepton-flavor-violating processes significantly and a few hundred GeV mass is possible.
\item The {two singly charged scalars} $h_1^\pm$ and $h_2^\pm$ have couplings to the charged leptons at the tree level.
Since in our model a part of couplings are sizable, constraints from lepton flavor violations and vacuum stability do not allow a few hundred GeV masses, especially {when} $k^{\pm\pm}$ is around a few hundred GeV.
The result of the global analysis in our previous paper~\cite{Nishiwaki:2015iqa} says that when $k^{\pm\pm}$ is $250\,\text{GeV}$ (which is around the minimum value of $m_{k^{\pm\pm}}$), $m_{h_1^\pm}$ and $m_{h_2^\pm}$ should be greater than $3\,\text{TeV}$.
\item In allowed parameter configurations, we found that the {absolute value of the} coupling $(y_R)_{22}$ (in front of $\bar{N}_{R_2} e^c_{R_2} h_2^-$), tends to be $8 \sim 9$, while the peak of the distribution of the scalar trilinear couplings $\mu_{11} \equiv \lambda_{11} v'/\sqrt{2}$ (in front of $h_1^- h_1^- k^{++}$) and $\mu_{22}$ (in front of $h_2^+ h_2^+ k^{--}$) is around $14 \sim 15\,\text{TeV}$.
We assumed that values of $\mu_{11}$ and $\mu_{22}$ are the same {and real} in the analysis.
\item The two $CP$ even components are mixed each other {as} shown in Eq.~(\ref{eq:CP-even_matrix}).
By the (simplified) global analysis in~\cite{Nishiwaki:2015iqa} based on the data~\cite{Aad:2014eha,Aad:2014eva,ATLAS:2014aga,Aad:2014xzb,ATLAS-CONF-2014-061,Khachatryan:2014jba}, the sine of the mixing angle $\alpha$ should be
\al{
{|\sin{\alpha}|} \lesssim 0.3,
	\label{eq:sinalpha_Higgssearch}
}
within $2\sigma$ allowed regions.
\item On the other hand, the observed relic density requires a specific range of $\sin{\alpha}$.
In our model, the Majorana DM $N_{R_1}$ communicates with the SM particles and the $U(1)$ NG boson $G$ through the two $CP$ even scalars $h$ and $H$.
When $v'$ is $\mathcal{O}(1)\,\text{TeV}$, DM\,--\,DM\,--\,$h/H$ couplings are significantly suppressed as $({M_{N_1}}/v')$ and then we should rely on the two scalar resonant regions.
When we consider the situation $m_{\text{DM}}/2 \simeq m_h\,(\simeq 125\,\text{GeV})$, a reasonable amount of the mixing angle $\alpha$ is required as
\al{
{|\sin{\alpha}|} \gtrsim 0.3,
}
where a {tense situation} with Eq.~(\ref{eq:sinalpha_Higgssearch}) is observed.
The allowed range of $v'$ is a function of $\sin{\alpha}$ and the maximum value is
\al{
v'|_{\text{max}} \sim 9\,\text{TeV} \ \text{when} \  {|\sin{\alpha}|} \sim 0.3.
}
Whereas the other resonant point is selected as $m_{\text{DM}}/2 \simeq m_H$, the requirement on the angle is as follows,
\al{
{|\sin{\alpha}|} \lesssim 0.3,
}
when $m_H = 250\,\text{GeV}$ or a bit more.
{We find that the heavy $H$ as $m_H = 500\,\text{GeV}$} cannot explain the relic density because of the suppression in the resonant propagator of $H$.
The maximum of $v'$ is found as
\al{
v'|_{\text{max}} \sim 6\,\text{TeV} \ \text{when} \  0 \lesssim {|\sin{\alpha}|} \lesssim 0.05,
	\label{eq:vmax_Hresonance}
}
where {the couplings of $H$} to the SM particles becomes so week and hard to be excluded from the 8\,TeV LHC results.
\end{enumerate}

\section{Decay widths at one loop
\label{sec:app:width}}

Here, we summarize the forms of relevant decay widths at one-loop level in the presence of the scalar mixing in Eq.~(\ref{eq:scalar_mixing_formulation}).
We mention that we ignore $\Gamma_{S' \to gg}$ since this value is tiny because of the fact $\B(H_{750\,\text{GeV}}^{\text{SM}} \to gg) = 2.55 \times 10^{-2}\%$.
The widths of the $125\,\text{GeV}$ Higgs boson are used for global fits of signal strengths of the observed Higgs.
\al{
\Gamma_{h \to gg}
&=
\frac{\alpha_{s}^2 m_h^3}{72 \pi^3 v^{2}} \left|\frac{3}{4} \left( A_{1/2}^{\gamma\gamma}(\tau_t^{\text{SM}}) \right) c_{\beta} \right|^2, \\
\Gamma_{h \to \gamma\gamma}
&=
\frac{\alpha_{\text{EM}}^2 m_h^3}{256\pi^3 v^{2}} \left|\left( A_1^{\gamma\gamma}(\tau_W^{\text{SM}}) + N_C Q_t^2 A_{1/2}^{\gamma\gamma}(\tau_t^{\text{SM}}) \right) c_{\beta} + \frac{1}{2} \frac{v [\sum_{a} \mu_{a}]}{m_{k^{\pm\pm}}^2} Q_k^2 A_0^{\gamma\gamma}(\tau_k^{\text{SM}}) (-s_{\beta}) \right|^{2}, \\
\Gamma_{h \to Z\gamma}
&=
\frac{{\alpha_{\text{EM}}^2} m_{h}^3}{512\pi^3} \left( 1 - \frac{m_Z^2}{m_h^2} \right)^3
\left| {A_{\text{SM}}^{Z\gamma}} c_{\beta} - \frac{ [\sum_{a} \mu_{a}] }{m_{k^{\pm\pm}}^2} \left( 2 Q_k g_{Zkk} \right) A_0^{Z\gamma}(\tau_k^{\text{SM}},\lambda_k) (-s_{\beta}) \right|^2, \\
\Gamma_{S' \to \gamma\gamma}
&=
\frac{\alpha_{\text{EM}}^2 m_{S'}^3}{256 \pi^3 v^{2}} \left|\left( A_1^{\gamma\gamma}(\tau_W) + N_C Q_t^2 A_{1/2}^{\gamma\gamma}(\tau_t) \right) s_{\beta} + \frac{1}{2} \frac{v [\sum_{a} \mu_{a}]}{m_{k^{\pm\pm}}^2} Q_k^2  A_0^{\gamma\gamma}(\tau_k) \, c_{\beta} \right|^{2}, \\
\Gamma_{S' \to Z\gamma}
&=
\frac{{\alpha_{\text{EM}}^2} m_{S'}^3}{512\pi^3} \left( 1 - \frac{m_Z^2}{m_{S'}^2} \right)^3
\left| {A_{\text{SM}}^{Z\gamma}}(\tau_{W,t}^{\text{SM}} \to \tau_{W,t}) s_{\beta} - \frac{[\sum_{a} \mu_{a}]}{m_{k^{\pm\pm}}^2} \left( 2 Q_k g_{Zkk} \right) A_0^{Z\gamma}(\tau_k,\lambda_k) \, c_{\beta} \right|^2, \\
\Gamma_{S' \to ZZ}
&=
\left| \left( \Gamma_{\text{tot}}(H^{\text{SM}}_{750\,\text{GeV}}) \B(H^{\text{SM}}_{750\,\text{GeV}} \to ZZ) \right)^{1/2} \, s_{\beta} + \mathcal{M}_{S \to ZZ} \, c_{\beta}  \right|^2,
}
with the factors
\al{
A_{\text{SM}}^{Z\gamma} &= \frac{2}{v} \left[ \cot{\theta_W} A_{1}^{Z\gamma}(\tau_W^{\text{SM}},\lambda_W) + N_C
\frac{(2Q_t)(T_3^{(t)} - 2Q_t s_W^2)}{s_W c_W} A_{{1/2}}^{Z\gamma}(\tau_t^{\text{SM}},\lambda_t) \right], \\
\mathcal{M}_{S \to ZZ} &= 
\left\{
\left( \frac{s_{W}^{2}}{2 c_{W}^{2}} \right)
\frac{{\alpha_{\text{EM}}^2} m_{S'}^3}{512\pi^3} \left( 1 - \frac{m_Z^2}{m_{S'}^2} \right)^3
\right\}^{1/2}
\left[
- \frac{[\sum_{a} \mu_{a}]}{m_{k^{\pm\pm}}^2} \left( 2 Q_k g_{Zkk} \right) A_0^{Z\gamma}(\tau_k,\lambda_k)
\right], \\[4pt]
A_1^{\gamma\gamma}(x)
&=
-x^2 \left[ 2x^{-2} + 3x^{-1} + 3(2x^{-1}-1) f(x^{-1}) \right], \\
A_{1/2}^{\gamma\gamma}(x)
&=
2x^2 \left[ x^{-1} + (x^{-1} -1) f(x^{-1}) \right], \\
A_{1}^{Z\gamma}(x,y)
&=
4 (3 - \tan^2{\theta_W}) I_2(x,y) + \left[ (1+2x^{-1}) \tan^2{\theta_W} - (5+2x^{-1}) \right] I_1(x,y), \\
A_{1/2}^{Z\gamma}(x,y)
&=
I_1(x,y) - I_2(x,y).
}

\noindent
Here, the ratios and two functions are defined for convenience
\al{
\tau_i^{\text{SM}} &= \frac{4 m_i^2}{m_h^2}, \quad
\tau_i              = \frac{4 m_i^2}{m_{S'}^2} \quad (i=t,W,k), \\
I_1(x,y) &= \frac{xy}{2(x-y)} + \frac{x^2y^2}{2(x-y)^2} \left[ f(x^{-1}) - f(y^{-1}) \right] + \frac{x^2y}{(x-y)^2} \left[ g(x^{-1}) - g(y^{-1}) \right] = A_{0}^{Z\gamma}(x,y), \\
I_2(x,y) &= - \frac{xy}{2(x-y)} \left[ f(x^{-1}) - f(y^{-1}) \right].
}
$\alpha_{s}$, $N_C\,(=3)$, $Q_t\,(=2/3)$ and $T_3^{(t)}\,(=1/2)$ are the fine structure constant of the QCD coupling, the QCD color factor for quarks, the electric charges of the top quark in unit of the positron's one, and the weak isospin of the top quark, respectively.
Other variables were already defined around Eqs.~(\ref{effective_couplings_oneloopHiggs})-(\ref{g(z)form}).
When we take the limit $s_{\beta} \to 0$, $\Gamma_{S' \to ZZ}$ is reduced to the form in Eq.~(\ref{eq:StoVV_1}).

\section{Additional plots
\label{sec:app:plots}}

\begin{figure}[t]
\begin{center}
\includegraphics[width=0.49\columnwidth]{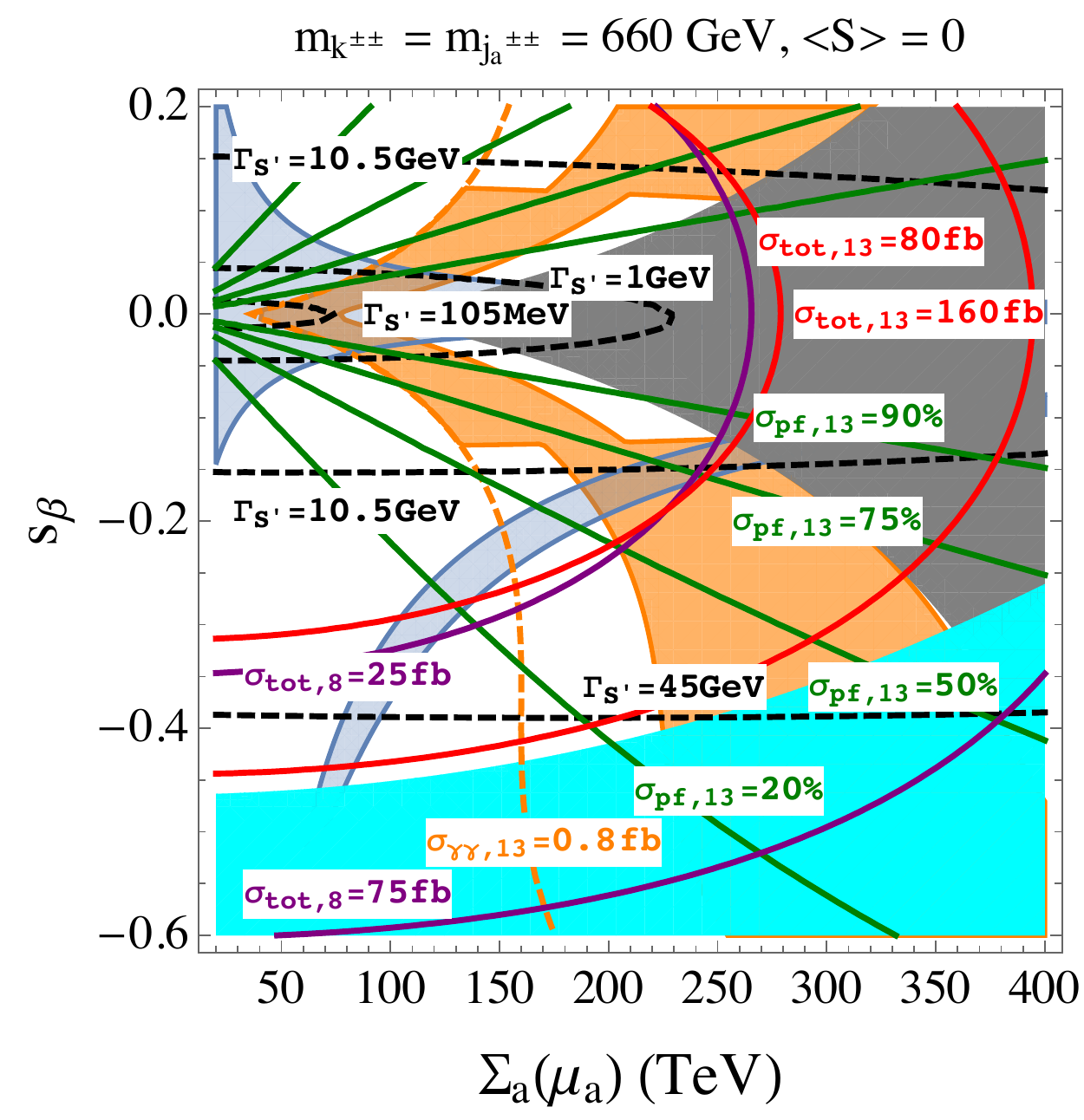}\
\includegraphics[width=0.49\columnwidth]{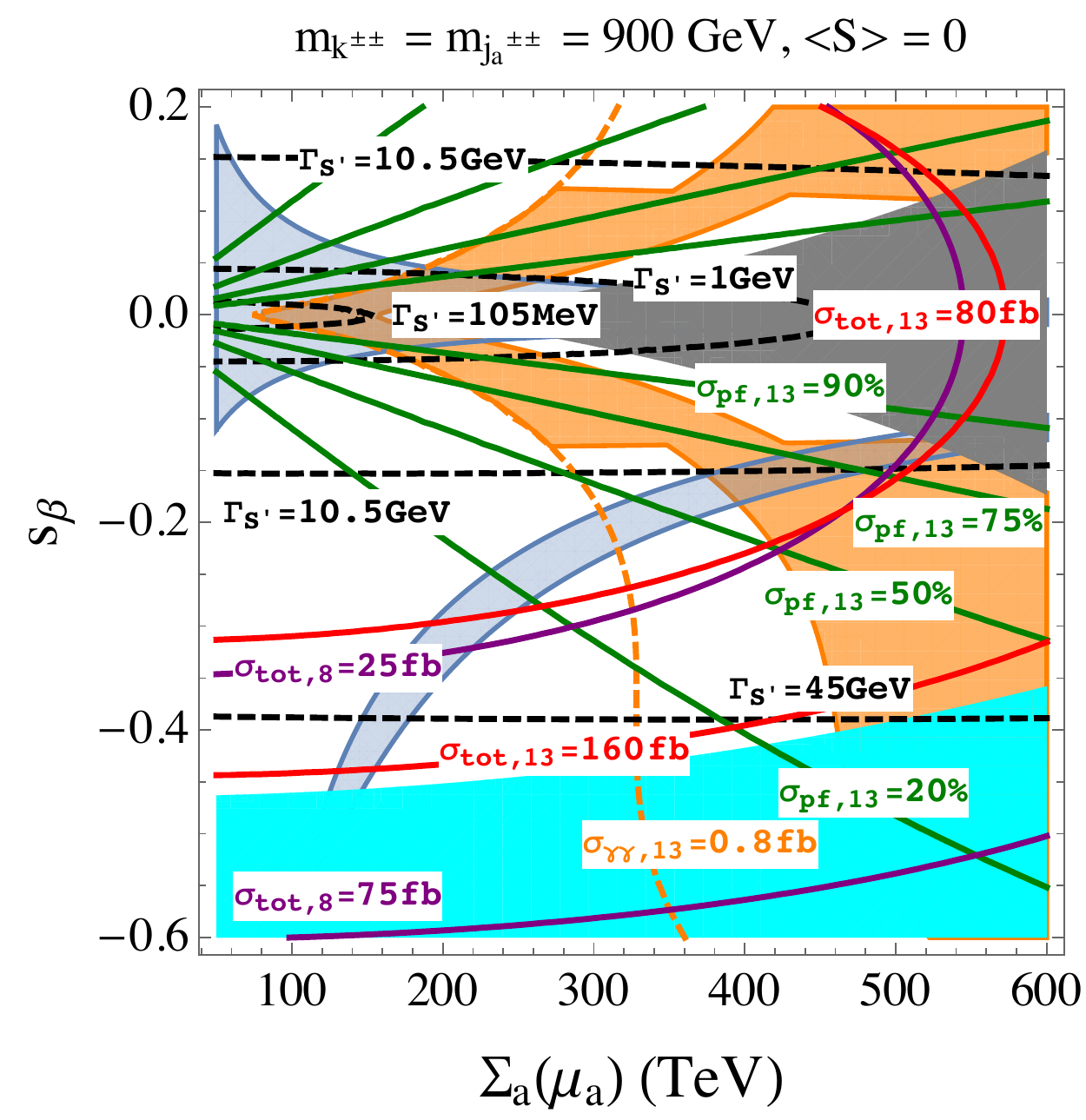}\\[4pt]
\caption{Allowed ranges of the parameters $\{ \sum_{a}(\mu_{a}), s_{\beta} \}$ are shown in the two choices of the mass of the degenerated {doubly charged} scalars ($m_{k^{\pm\pm}} \, [= m_{j_{a}^{\pm\pm}}] = 660/900\,\text{GeV}$ {[left panel/right panel]}).
Under the assumption $\langle S \rangle = 0$, the value of $m_{S'h}$ is fixed as shown in Eq.~(\ref{eq:m_correspondence}).
The light blue regions represent $2\sigma$ allowed regions of $125\,\text{GeV}$ Higgs signal strengths, while the orange regions suggest the areas where the $750\,\text{GeV}$ excess is suitably explained.
The gray/cyan regions are excluded in $95\%$ C.L.s by the ATLAS $8\,\text{TeV}$ results for $S' \to \gamma\gamma/ZZ$.
For better understanding, several contours for the total width of $S'$ ($\Gamma_{S'}$),
total production cross sections at $\sqrt{s} = 8/13\,\text{TeV}$ ($\sigma_{\text{tot},8/13}$),
and the percentage of the production through the photon fusion at $\sqrt{s} = 13\,\text{TeV}$ ($\sigma_{\text{pf},13}$) are illustrated. 
}
   \label{fig:allowedregion_2}
\end{center}
\end{figure}

In this appendix, we provide plots for discussing the case that the mixing of two fields $S$ and $\Phi$ through mass terms under the assumption $\langle S \rangle = 0$.
Here, the mass parameter $m_{S' h}$ in the $S'$-$h$-$h$ interaction is automatically determined by the two mass
eigenvalues and the mixing angle $\beta$ as shown in Eq.~(\ref{eq:m_correspondence}).
We note that the two choices in the universal mass of {doubly charged} scalars ($660\,\text{GeV}$ and $900\,\text{GeV}$) are from the expected $95\%$ C.L. lower bounds under the assumption $\B(j_{a}^{\pm\pm} \to \mu^{\pm}\mu^{\pm}) = 100\%$ when $N_{j} = 10$ and $N_{j} = 100$, respectively.

\begin{figure}[t]
\begin{center}
\includegraphics[width=0.45\columnwidth]{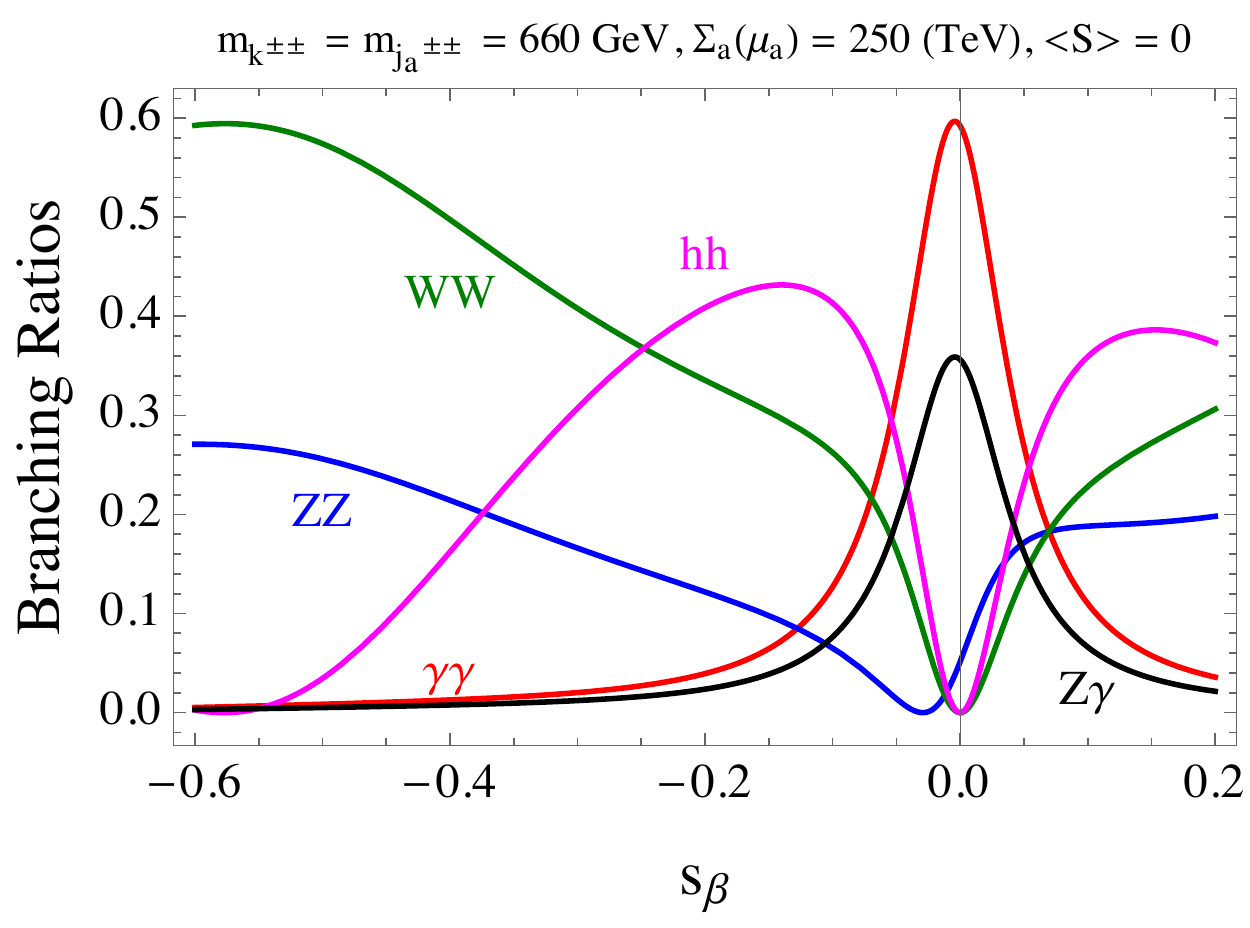}\
\includegraphics[width=0.45\columnwidth]{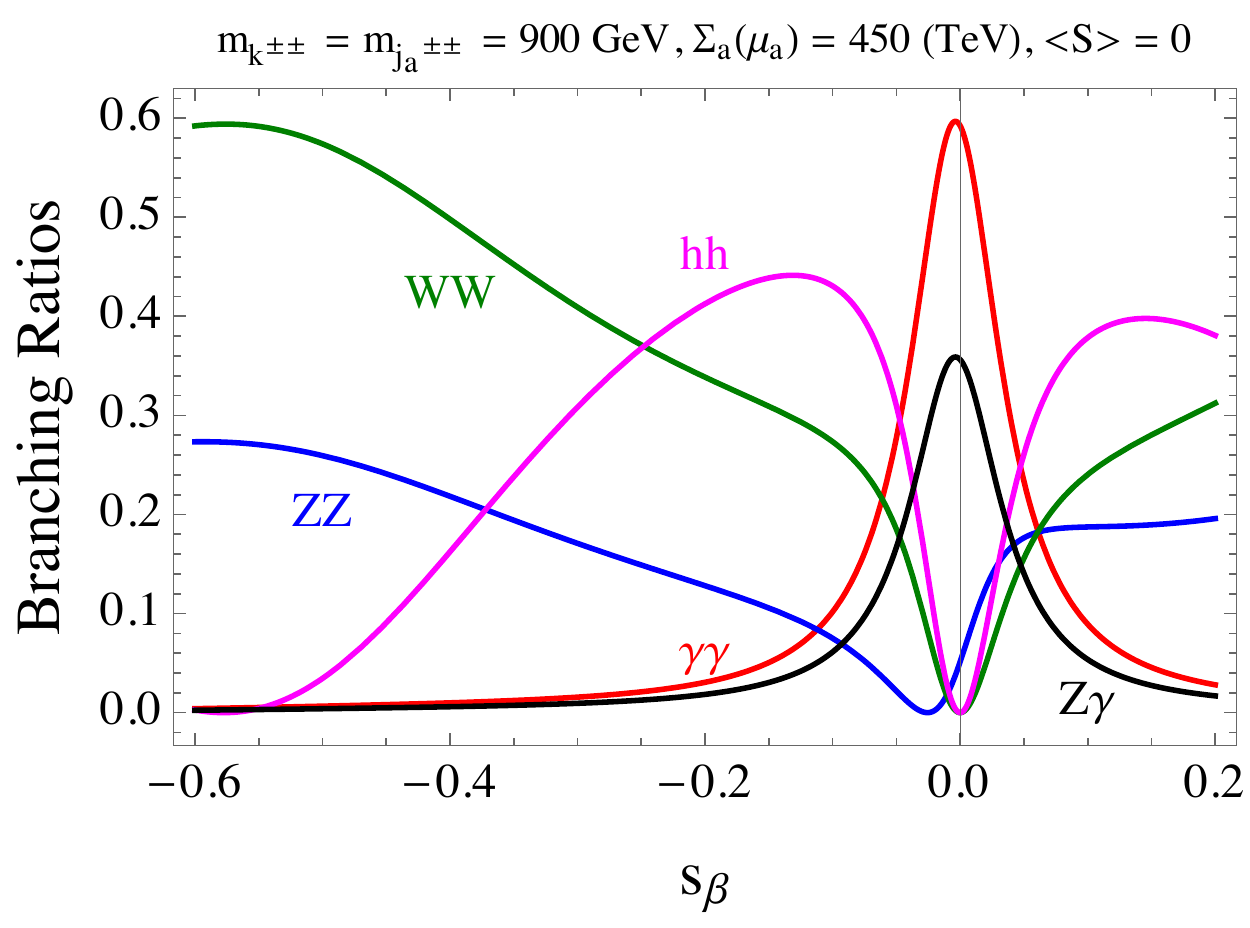}\\[4pt]
\caption{Relevant branching ratios of $S'$ in the {two} configurations in Fig.~\ref{fig:allowedregion_2} are shown.
Here, values of $\Sigma_{a} (\mu_{a})$ are suitably fixed as typical digits in the corresponding allowed regions.
}
   \label{fig:Sprime_Brs_2}
\end{center}
\end{figure}


\bibliographystyle{utphys}
\bibliography{draft_750GeV}

\end{document}